\documentclass[sigconf]{acmart}

\AtBeginDocument{%
  \providecommand\BibTeX{{%
    \normalfont B\kern-0.5em{\scshape i\kern-0.25em b}\kern-0.8em\TeX}}}

\usepackage{multirow}
\usepackage{makecell}
\usepackage{color, colortbl}

\usepackage{comment}
\usepackage{xspace}
\usepackage{array}
\usepackage{subcaption}

\newcommand{\technique}[1]{TECHNIQUE}
\newcommand{\system}[1]{\textsc{PriorWeaver\xspace}}

\newcommand{\eunice}[1]{{\color{magenta} Eunice: #1 }}

\newcommand{\revision}[1]{{#1}}
\newcommand{\secondRevision}[1]{{#1}}

\newcommand{\userQuote}[1]{\textit{``{#1}''}}

\copyrightyear{2026}
\acmYear{2026}
\setcopyright{cc}
\setcctype{by}
\acmConference[CHI '26]{Proceedings of the 2026 CHI Conference on Human Factors in Computing Systems}{April 13--17, 2026}{Barcelona, Spain}
\acmBooktitle{Proceedings of the 2026 CHI Conference on Human Factors in Computing Systems (CHI '26), April 13--17, 2026, Barcelona, Spain}
\acmDOI{10.1145/3772318.3790349}
\acmISBN{979-8-4007-2278-3/2026/04}

\begin{document}

\title{\system{}: Prior Elicitation via Iterative Dataset Construction}

\author{Yuwei Xiao}
\email{yuweix@ucla.edu}
\orcid{0009-0005-1754-6565}
\affiliation{%
  \institution{UCLA}
  \city{Los Angeles}
  \state{California}
  \country{USA}
}

\author{Shuai Ma}
\email{mashuai@iscas.ac.cn}
\orcid{0000-0002-7658-292X}
\affiliation{%
  \institution{Aalto University}
  \city{Helsinki}
  \country{Finland}
}

\author{Antti Oulasvirta}
\email{antti.oulasvirta@aalto.fi}
\orcid{0000-0002-2498-7837}
\affiliation{%
  \institution{Aalto University}
  \city{Helsinki}
  \country{Finland}
}

\author{Eunice Jun}
\email{emjun@cs.ucla.edu}
\orcid{0000-0002-4050-4284}
\affiliation{%
    \institution{UCLA}
    \city{Los Angeles}
    \state{California}
    \country{USA}
}

\renewcommand{\shortauthors}{Xiao et al.}

\begin{abstract}

In Bayesian analysis, prior elicitation, or the process of \revision{facilitating the expression of} one’s beliefs to inform statistical modeling, is an essential yet challenging step. Analysts often have beliefs about real-world variables and their relationships. However, existing tools require analysts to translate these beliefs and express them indirectly as probability distributions over model parameters. We present \system{}, an interactive visualization system that facilitates prior elicitation through iterative dataset construction and refinement. Analysts visually express their assumptions about individual variables and their relationships. Under the hood, these assumptions create a dataset used to derive statistical priors. Prior predictive checks then help analysts compare the priors to their assumptions. In a lab study with 17 participants new to Bayesian analysis, we compare \system{} to a baseline incorporating existing techniques. Compared to the baseline, \system{} gave participants greater control, clarity, and confidence, leading to priors that were better aligned with their expectations.

\end{abstract}

\ccsdesc[500]{Human-centered computing~Interactive systems and tools}
\ccsdesc[500]{Human-centered computing~Visualization systems and tools}

\keywords{Bayesian statistics, statistical analysis, prior elicitation, belief elicitation, interactive visualization, dataset construction}


\begin{teaserfigure}
  \centering
  \includegraphics[width=0.95\linewidth]{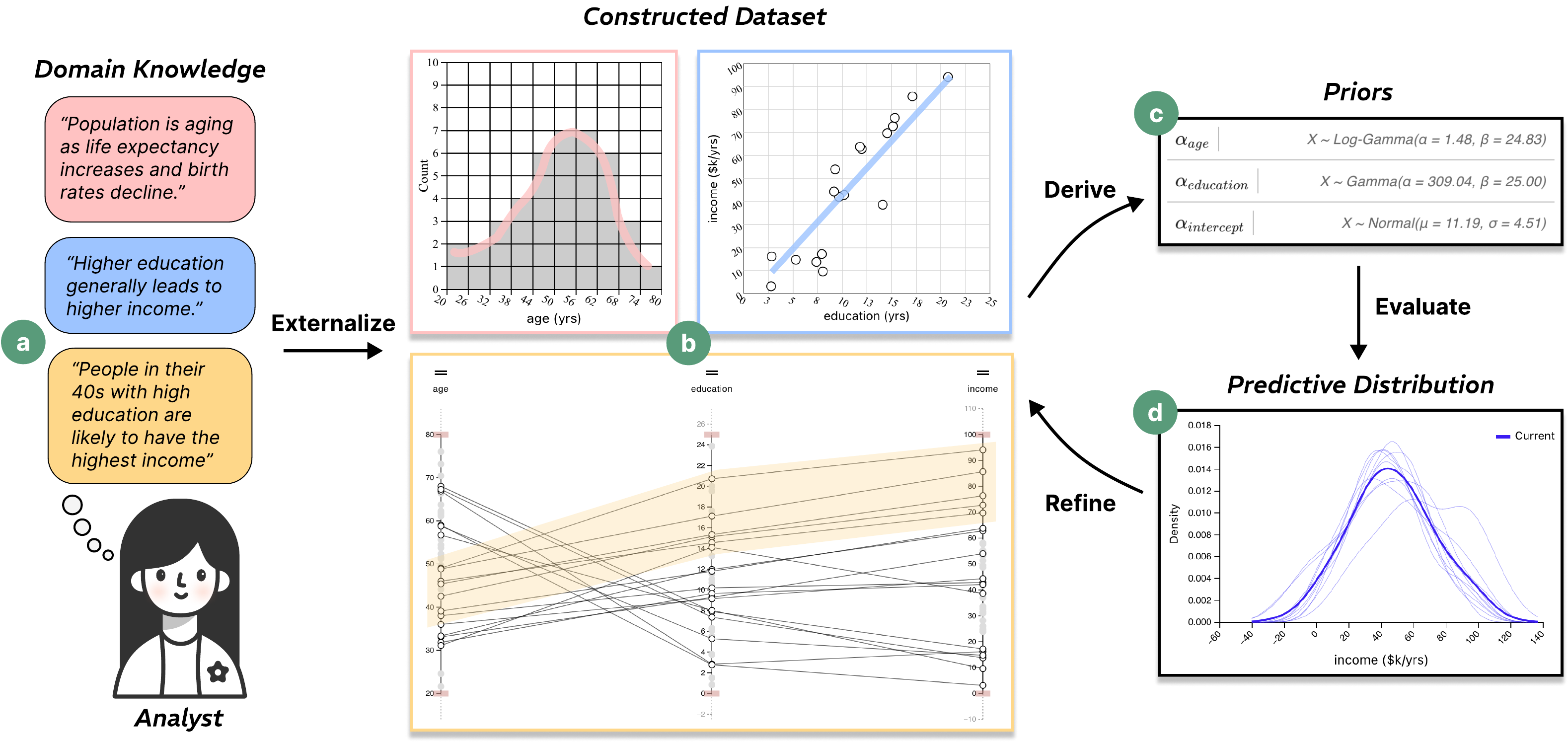}
  \caption{
  \system{} supports prior elicitation and iteration through the construction of a dataset representative of analysts' beliefs. (a) Analysts begin by considering their implicit domain knowledge about variable distributions (red), pairwise relationships (blue), and multivariate relationships (yellow). (b) Analysts express these assumptions through coordinated interactive visualizations, which simultaneously construct a representative dataset. (c) The dataset is used to derive statistical priors by fitting a predefined model. (d) Prior predictive checks visualize the predicted distribution of the outcome variable, which  analysts can compare to their assumptions and use to iterate on their inputs.
  }
  \label{fig:teaser}
  \Description[Diagram of the \system{} elicitation and iteration process.]{A four-stage flowchart: (a) Externalizing implicit domain knowledge; (b) Interactive visualization for dataset construction; (c) Deriving priors via model fitting; (d) Prior predictive checks for evaluation and iteration.}
\end{teaserfigure}

\maketitle

%

\section{Introduction}

\revision{
\emph{Bayesian analysis}, or Bayesian inference, is an approach to analyzing and learning from data. Bayesian analysis involves using Bayes' rule to update existing assumptions, or beliefs, about a domain based on new data, fit according to a specified statistical model\footnote{We refer readers interested in a more thorough introduction to Bayesian analysis to the summary provided by Phelan et al.~\cite{phelan-2019-template} and Richard McElreath's textbook \textit{Statistical Rethinking}~\cite{mcelreath2020statisticalRethinking}.}. 
Bayesian analysis is an alternative to the currently pervasive Frequentist approach, which does not explicitly incorporate existing domain assumptions into the analysis procedure. 
As a result, Bayesian analysis provides greater interpretability and more accurate accumulation of knowledge compared to Frequentist approaches~\cite{ioannidis2005most, phelan-2019-template}. Additionally, results from Bayesian analyses are easier to understand because they align with common-sense interpretations~\cite{oakes1986statistical, phelan-2019-template}. For these reasons and more, researchers have been advocating for the adoption of Bayesian analysis methods across disciplines~\cite{dienes2011bayesian,howard2016proof,kruschke2010bayesian}, including in HCI~\cite{kaptein2012rethinking,kay2016researcherStatistics}.
Kay et al. argue that Bayesian analysis is well-suited to HCI because it enables learning from small-sample studies, stabilizes estimates of effects, reduces overfitting to limited observations, and builds knowledge across studies accretively~\cite{kay2016researcherStatistics}.
}

Many of the aforementioned benefits stem from a key aspect of Bayesian analysis: the integration of analysts' domain knowledge and assumptions (e.g., \autoref{fig:teaser}a). 
Analysts imbue Bayesian analysis with their assumptions by specifying prior distributions for model parameters, or simply \emph{priors} (e.g., \autoref{fig:teaser}d).
The process of \revision{transforming domain knowledge into these probability distributions}, referred to as \emph{prior elicitation}, is critical yet difficult to do well~\cite{mikkola2024prior,ohagan-2006-uncertain}. Prior elicitation requires not only expertise about a domain (e.g., \textit{``Education year has a moderate positive impact to income''}) but also the ability to express that knowledge using mathematical distributions over model parameters (e.g., \textit{``The coefficient of education, given age, on income has a Normal(3, 0.5) distribution''}), which often do not have direct real-world analogs.
When priors fail to capture analysts’ domain knowledge, the subsequent inferences may be misleading or implausible, particularly in small-sample settings~\cite{Gelman2017likelihood}.
Indeed, a common practice is for domain experts to hire a Bayesian modeling expert, often referred to as a facilitator, to assist with prior elicitation, a dependency that poses a barrier to wider adoption of Bayesian analysis.

Over the years, prior elicitation tools have made progress, from systems that rely on knowledge over parameters~\cite{gosling-2018-shelf, morris2014match, preliz}, to approaches that let analysts reason more directly about outcomes~\cite{hartmann2020prior, bockting2024simulation}. 
Yet, gaps persist: most tools still rely on probability-format input, overlook relationships across variables, offer limited feedback for refinement, and provide little support for Bayesian novices.
%


We present \system{}, an interactive tool that guides analysts through prior elicitation. 
The key idea of \system{} is to approach prior elicitation as a dataset construction problem.
In \system{}, analysts can directly express their assumptions about possible values one could observe about variables in the real-world (e.g., \textit{``People in their 40s earn between \$40k and \$60k''}). 
This provides analysts a concrete and tangible representation of their domain knowledge.
%
%
%
Under the hood, \system{} constructs a concrete dataset from these inputs then derives prior distributions by fitting the statistical model to the dataset. 
\system{} outputs prior predictive check visualizations that help analysts compare predictive outcomes and refine these priors. \autoref{fig:teaser} gives an overview of this interactive process. 
%

%

To evaluate \system{}, we conducted a controlled within-subjects experiment. Seventeen analysts experienced in statistical modeling but new to Bayesian analysis (i.e., Bayesian novices) specified priors for a statistical model using both \system{} and a baseline parameter-based prior elicitation interface.
%
%
Compared to the baseline, analysts reported feeling that they could express their knowledge more comfortably and clearly using \system{}. 
They also produced initial priors that aligned closely with their beliefs and final priors. 
Analysts also reported that \system{} made the feedback more actionable, enabling more effective and purposeful refinement.
Our results suggest that shifting prior elicitation towards a constructive sensemaking process makes Bayesian analysis more approachable.  

This paper contributes:  

\begin{itemize}
    \item A new perspective on prior elicitation as the process of constructing a dataset that captures analysts' assumptions without requiring direct parameter specification;
    \item \system{}, an interactive system that supports iterative dataset construction through coordinated visualizations, derives statistical priors via bootstrapping, and provides feedback through prior predictive checks; and
    \item Evidence from a controlled lab study that \system{} gives analysts helpful structure for externalizing domain knowledge, control over refining priors, and more positive attitudes toward Bayesian analysis.
\end{itemize}  

\section{Background and Related Work}
\label{sec:related}


Our work builds on and contributes to the literature on interactive tools for prior elicitation, perspectives on what makes a good prior, and the role of visualization in belief elicitation.

\newcolumntype{C}[1]{>{\centering\arraybackslash}p{#1}}

\begin{table*}[tp]
\centering
\caption{\textbf{Comparison of \system{} with existing prior elicitation tools across key dimensions.}
Elicitation space refers to what knowledge is elicited, either about parameter or observable. 
Elicitation modality, or how an analyst inputs their priors, is either graphical or textual. 
Elicitation format refers to the type of input required from the analysts. \textit{Probability} denotes probabilistic input (e.g., mean, variance), while \textit{Samples} denotes hypothetical samples. 
Feedback mechanism is how a tool shows analysts their priors. \textit{PPCs} refers to prior predictive checks, a common approach in Bayesian analysis. 
Multivariate refers to elicit beliefs about multiple parameters or variables simultaneously.
\revision{Some tools support both categories within a dimension; for conciseness, we list only the primary category they support and mark it with an asterisk (*).} 
Overall, existing tools typically operate in the parameter space, require probabilistic inputs, lack feedback mechanisms, and provide limited support for eliciting multiple parameters or variables simultaneously. In contrast, \system{} integrates and extends these dimensions to offer more comprehensive support.
}
{\renewcommand{\arraystretch}{1.4}
\begin{tabular}{l C{2cm} C{2cm} C{2cm} C{2.2cm} C{1.8cm}}
\toprule
\textbf{Tool}
& \textbf{Elicitation Space} 
& \textbf{Elicitation Modality} 
& \textbf{Elicitation Format} 
& \textbf{Feedback \newline Mechanism} 
& \textbf{Multivariate} \newline ($N > 2$) \\
\midrule
Jones and Johnson~\cite{jones2014prior} & Parameter & Textual & Probability & -- & $\times$ \\
\textsc{match}~\cite{morris2014match} & Parameter & Graphical* & Probability & -- & $\times$ \\
\textsc{shelf}~\cite{gosling-2018-shelf} & Parameter & Graphical* & Probability & -- & $\times$ \\
Sarma and Kay~\cite{sarma2020prior} & Parameter & Graphical & Probability & PPCs & $\times$ \\
\textsc{preliz}~\cite{preliz} & Parameter & Graphical* & Probability & PPCs & $\checkmark$ \\
Hartmann et al.~\cite{hartmann2020prior} & Observable & Textual & Probability & -- & $\checkmark$ \\
Bockting et al.~\cite{bockting2024simulation} & Observable* & Graphical* & Probability & -- & $\checkmark$ \\
Casement et al.~\cite{casement2018graphical} & Observable & Graphical & Samples & -- & $\times$ \\
\textbf{\system{}} & \textbf{Observable} & \textbf{Graphical} & \textbf{Samples} & \textbf{PPCs} & $\mathbf{\checkmark}$ \\
\bottomrule
\end{tabular}}
\label{tab:design-space}
\vspace{-\baselineskip}
\end{table*}


\subsection{Tools for Prior Elicitation}
\label{sec:related:tool}

The majority of prior elicitation tools operate in the \textbf{\textit{parameter space}}, requiring analysts to express knowledge directly about statistical parameters of the model. 
For example, \textsc{shelf}~\cite{gosling-2018-shelf} and \textsc{match}~\cite{morris2014match} support this process by asking analysts to specify moments (e.g., mean, variance), quantiles, or histograms \revision{(e.g., trial-roulette method~\cite{gore1987biostatistics}, which lets analysts distribute probability mass across bins)}, and then fitting a distribution according to these inputs. 
\revision{Additionally, the \textsc{preliz} tool, the system from Jones and Johnson~\cite{jones2014prior}, and Sarma and Kay's tool~\cite{sarma2020prior} utilize prior predictive checks (PPCs)~\cite{gabry2019visualization}. PPCs draw samples from the model’s prior distributions and use them to simulate the kinds of data the model would produce, also known as predictive distributions. This enables analysts to engage in \emph{predictive exploration}~\cite{kadane1998experiences}, a process of changing priors and assessing whether the chosen priors generate plausible data that aligns with analysts' domain knowledge.}
%
%
While predictive exploration improves interpretability, these tools still require analysts to reason about abstract statistical parameters whose behavior and real-world implications are often difficult to intuit.

An alternative is to elicit priors in the \textbf{\textit{observable space}}, where analysts reason directly about measurable quantities like model variables. This aligns more closely with domain expertise, allowing analysts to specify patterns they have observed without translating them into parameters~\cite{ohagan-2006-uncertain, mikkola2024prior}. For example, Casement et al.~\cite{casement2018graphical} ask analysts to iteratively select the most plausible histogram from a set of simulated samples. Then, their approach infers a prior from these selections. However, this method is limited to univariate models. 
More recent simulation-based approaches remove this requirement by letting experts specify expectations about multiple observable quantities and then automatically searching for priors that produce matching predictions. Techniques such as multi-objective Bayesian optimization~\cite{manderson2023translating} and stochastic gradient-based optimization~\cite{hartmann2020prior, bockting2024simulation} learn priors that minimize the discrepancy between simulated and elicited information. 
\revision{However, these approaches still rely primarily on probabilistic inputs, fail to capture relational knowledge between variables, or provide limited feedback for validating priors. 
More importantly, these approaches have not yet been developed into fully functional tools, let alone as interfaces designed for end-users.}
In contrast, \system{} enables analysts to externalize their knowledge in the observable space as a tangible dataset through coordinated visualizations. Analysts can then iteratively validate and refine these priors with the help of prior predictive checks.

Table~\ref{tab:design-space} summarizes the characteristics of prior elicitation tools and compares \system{} against existing tools.
\revision{In short, existing prior elicitation tools are mainly designed for Bayesian practitioners who are familiar with the Bayesian workflow, rather than domain experts who wish to conduct Bayesian analysis but lack formal Bayesian training. } 
Moreover, there is little consensus on what abstractions (parameter-space vs. observable-space) or interfaces help analysts express their domain knowledge as priors at all. 
This work articulates and evaluates design considerations for facilitating domain knowledge expression for the purpose of eliciting priors.

\subsection{What Makes a Good Prior?}
\label{sec:related:prior}


What constitutes a good prior is highly contested within the Bayesian modeling community~\cite{berger2006case, gelman2017beyond}. The primary dimension along which priors differ is \emph{informativeness}, or how much influence a prior exerts on the model fitting process. This depends on how narrow (e.g., low variance) or broad (e.g., high variance) a distribution is, which directly influences how much the collected data are prioritized when fitting a statistical model 
(See Figure 1 in ~\cite{sarma2020prior}). 

At one extreme is a \emph{non-informative} prior, which has high variance and wide tails and, as a result, prioritizes the collected data during the model fitting process over an analyst's domain knowledge. At the other extreme is an \emph{informative} prior, which is more opinionated, has a smaller variance, and exerts strong influence on the model fitting process. In between these extremes lies the \emph{weakly informative} prior~\cite{Gelman2008weaklyprior}, which is intentionally specified to contain less information than what might be available. In practice, there is limited guidance about where the boundaries between these categories lie, how to use them, and why one is preferable over another in the same analysis setting.

Moreover, whether one type of prior distribution is better than another remains debated. Key considerations are whether priors lead to useful statistical predictions and to what extent they reflect the analyst’s implicit understanding of the domain. For example, a prior may support strong predictive performance but fail to capture how an analyst conceptualizes the problem. 
%


\system{}'s goal is to elicit priors that faithfully reflect analysts’ knowledge~\cite{garthwaite2005statistical}, which depend on analysts' beliefs and may span across the informativeness spectrum. \system{} prioritizes the specification of priors that would generate data consistent with the analyst’s understanding of the world~\cite{Gelman2017likelihood}.


\subsection{Effects of Visualization on Belief Elicitation}
\label{sec:related:belief}
\revision{Prior elicitation is a form of belief elicitation in which a researcher or system helps individuals externalize their assumptions.}
Research shows that analysts often ground their beliefs in conceptual models rather than data~\cite{klahr1988dual, choi2019concept, jun2022hypothesis}, and that making these beliefs explicit improves recall and reflection~\cite{kim2017gap, jun2024rtisane}.
Recent visualization experiments further demonstrate that users externalize more data assumptions when sketching than when verbalizing~\cite{koonchanok2021prophecy}.
%

Building on these findings, researchers propose various visualization techniques for belief elicitation.
\revision{Some have investigated how frequency framing can be applied to visualizations to facilitate eliciting belief.}
Goldstein and Rothschild demonstrate that drawing full distributions yields more accurate results than providing quantiles~\cite{goldstein2014understanding}. 
Other studies show that frequency formats often support better reasoning than probability formats~\cite{gigerenzer1995improve, kay2016ish, hullman2017imagining}.
Kim et al. introduce a graphical sample-based elicitation interface in which users provide individual sample values that are aggregated into a distribution~\cite{kim2019bayesian}. 
%
\revision{In addition, recent work have begun integrating interaction techniques into visualizations to better support belief elicitation.}
Karduni et al. propose the Line + Cone method for eliciting beliefs about bivariate correlation, where users draw a central trend and specify their uncertainty around it~\cite{karduni2020bayesian}.
Koonchanok et al. introduce an interactive scatterplot to elicit bivariate beliefs where users can adjust the slope of a trendline to indicate the expected relationship, and adjust the expected uncertainty in the relationship using a slider~\cite{koonchanok2023visualbelief}.
These techniques highlight diverse approaches to belief elicitation. Mahajan et al.~\cite{mahajan2022vibe} synthesize this body of work into VIBE, a design space that organizes belief-driven visualizations by elicitation goals and input modalities.

Extending belief visualization research that emphasizes comparing beliefs to data, \system{} anchors elicitation in the construction of a concrete, manipulable dataset that serves as the foundation for incorporating analysts' beliefs into mathematical models in Bayesian analysis.

\section{Design Considerations}
\label{sec:design}
Existing prior elicitation tools rarely focus on observable-space elicitation, often rely on probability-format input, overlook relational knowledge across variables, and offer limited feedback for refinement.
\revision{These gaps hinder analysts, especially domain experts new to Bayesian analysis, from incorporating their knowledge and specifying appropriate priors during Bayesian analysis.}
To address these gaps and incorporate suggestions in prior work~\cite{kadane1998experiences, garthwaite2005statistical, gelman2020workflow, mikkola2024prior, gabry2019visualization}, we articulated four design considerations that informed the development of \system{}:

\revision{
\textbf{DC1: Support knowledge expression in the observable space.}
}
Analysts are more familiar with observable quantities (e.g., values of age, income, education) than with abstract model parameters (e.g., regression coefficients)~\cite{kadane1998experiences, ohagan-2006-uncertain}. 
Analysts should be able to work with representations that map directly to their domain knowledge when specifying priors. 

\revision{
\textbf{DC2: Support expressions of both distributional and relational knowledge.}
}
Analysts have knowledge about not only distributions of individual variables but also relationships among variables (e.g., how education correlates with income)~\cite{kadane1998experiences}.
Yet existing observable-space tools often emphasize distributional specifications, while neglecting analysts' need to express relational knowledge~\cite{jun2024rtisane}. 
Systems should therefore enable analysts to express both distributions and relationships. 
Systems should also ensure these forms of knowledge are connected to capture the multidimensional nature of analysts' beliefs.

\revision{
\textbf{DC3: Offer actionable feedback to support evaluation and iterative refinement of priors.}
}
%
Frequent feedback is essential for enabling evaluation and iterative refinement of priors~\cite{kadane1998experiences}. 
While prior predictive checking~\cite{gabry2019visualization} is a common method for evaluating priors, the outputs (i.e., predictive distributions) are often disconnected from analysts’ inputs, offering little guidance for correction. 
Instead, systems should connect analysts' externalized knowledge with the resulting priors explicitly.

\revision{\textbf{DC4: Incorporate visual representations to facilitate the elicitation process.}}
Interactive visualizations provide interpretable ways to elicit inputs and evaluate priors~\cite{mikkola2024prior, gabry2019visualization}. 
As such, systems could incorporate multiple visualization types to help analysts express their beliefs more expressively and to support flexible elicitation strategies analysts may have~\cite{sarma2020prior}. 
In addition, frequency-based visualizations may be most effective given that prior work has found that people interpret and reason with frequency formats more effectively than probability formats~\cite{goldstein2014lay, hullman2017imagining, ohagan-2006-uncertain, ohagan2019knowledge}.

\section{\system{}: Prior Specification as Dataset Construction}
\label{sec:system}

\begin{figure*}[thbp]
	\centering 
	\includegraphics[width=\linewidth]{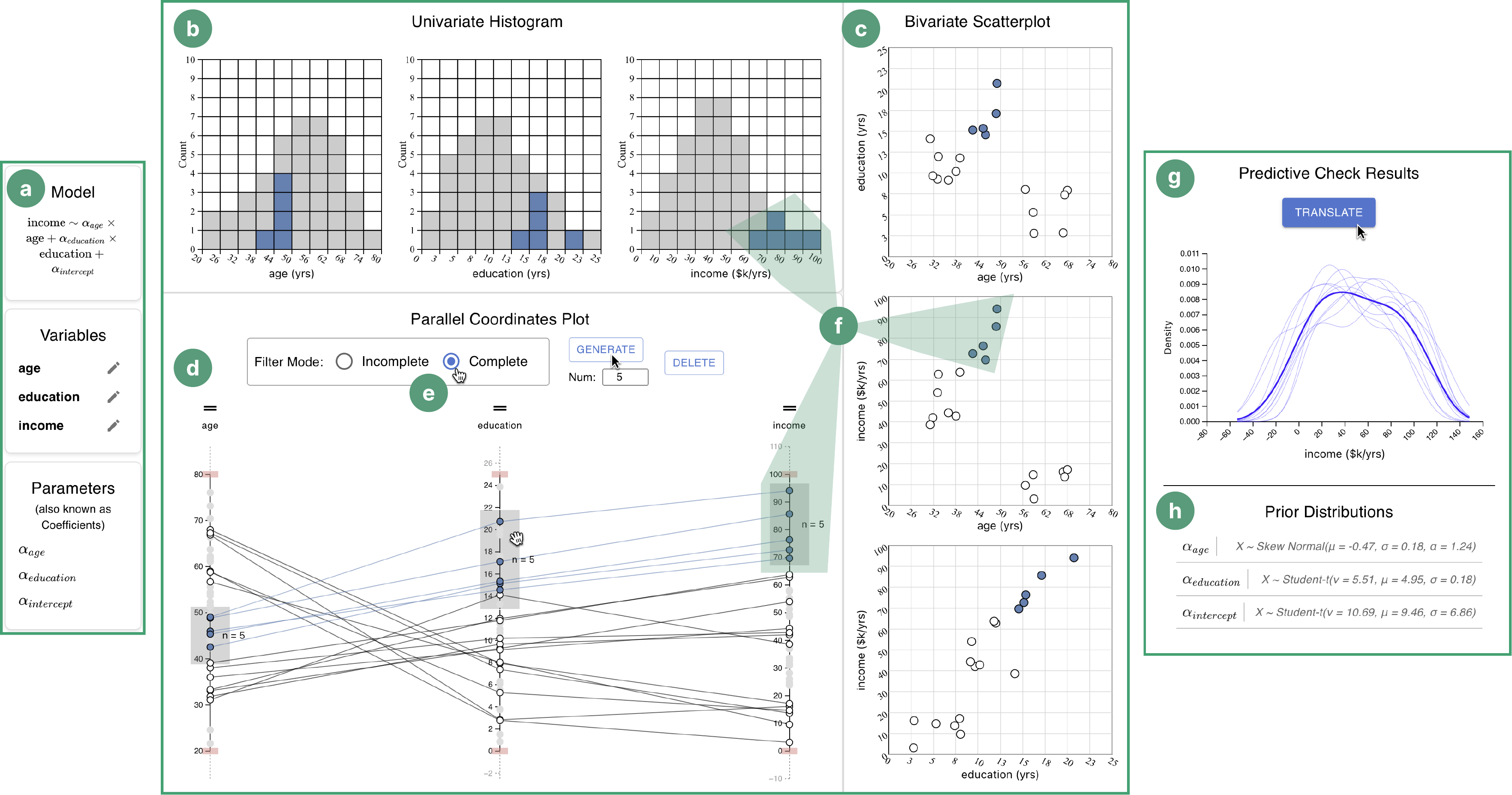}
	\caption{
    \textbf{
    \system{}'s user interface.}
    (a) An information panel displays the model formula, variables, and parameters.
    To externalize their knowledge for priors, analysts work in the central coordinated visualizations panel, which includes 
    (b) univariate histograms for variable distributions, 
    (c) bivariate scatterplots for pairwise relationships, 
    and (d) a parallel coordinates plot for multivariate relationships. 
    (f) Brushing on the parallel coordinates plots' axes serves as a cross-filter, with selections (blue dots) synchronized across all visualizations.
    (e) Analysts can toggle between displaying complete or incomplete entities (white dots), and hide the others (gray dots). In \textit{Complete} mode, they can use the \textsc{generate} function to define multivariate assumptions within brushed regions and add the generated entities to the constructed dataset.
    To derive and evaluate the priors, analysts can click \textsc{translate} to view (g) prior predictive checks and (h) suggested prior distributions.
    }
	\label{fig:ui}
    \Description[Screenshot of the \system{} user interface showing coordinated views.]{A multi-panel interface featuring: (a) Model info; (b) Univariate histograms; (c) Bivariate scatterplots; (d) Parallel coordinates plot; (f) Brushing-and-linking highlighting blue dots; (g) Predictive check results; and (h) Derived prior distributions.}
\end{figure*}

The central contribution of \system{} is the perspective that prior elicitation can be understood as a dataset construction problem. This perspective mirrors real-world practices, where analysts often reason about priors by referencing published datasets or concrete examples in mind~\cite{gelman2013philosophy, viele2014use, ohagan2019knowledge}. 
An analyst-constructed dataset serves as a tangible artifact of the analyst's knowledge: columns represent distributional knowledge of individual variables, while rows capture relational knowledge across variables. 
Through constructing this dataset, analysts externalize implicit assumptions as concrete values in the observable space. 
This conceptual reframing comes with three technical challenges: (i) how to support analysts in effectively constructing the dataset, (ii) how to derive statistical priors from the constructed dataset, and (iii) how to ensure that the derived priors accurately reflect analysts’ beliefs.


Below, we describe the system's design and implementation, including how \system{} addresses these three technical challenges.
%
%
\system{} currently supports prior elicitation for generalized linear models (GLMs) involving only continuous variables. We discuss challenges and opportunities for expanding support for more statistical models and variable types in Section~\ref{sec:limitations}.

\subsection{Externalizing Domain Assumptions via Coordinated Interactive Visualizations} \label{system:visualization}

\begin{figure*}[t!]
    \centering
    \includegraphics[width=\linewidth]{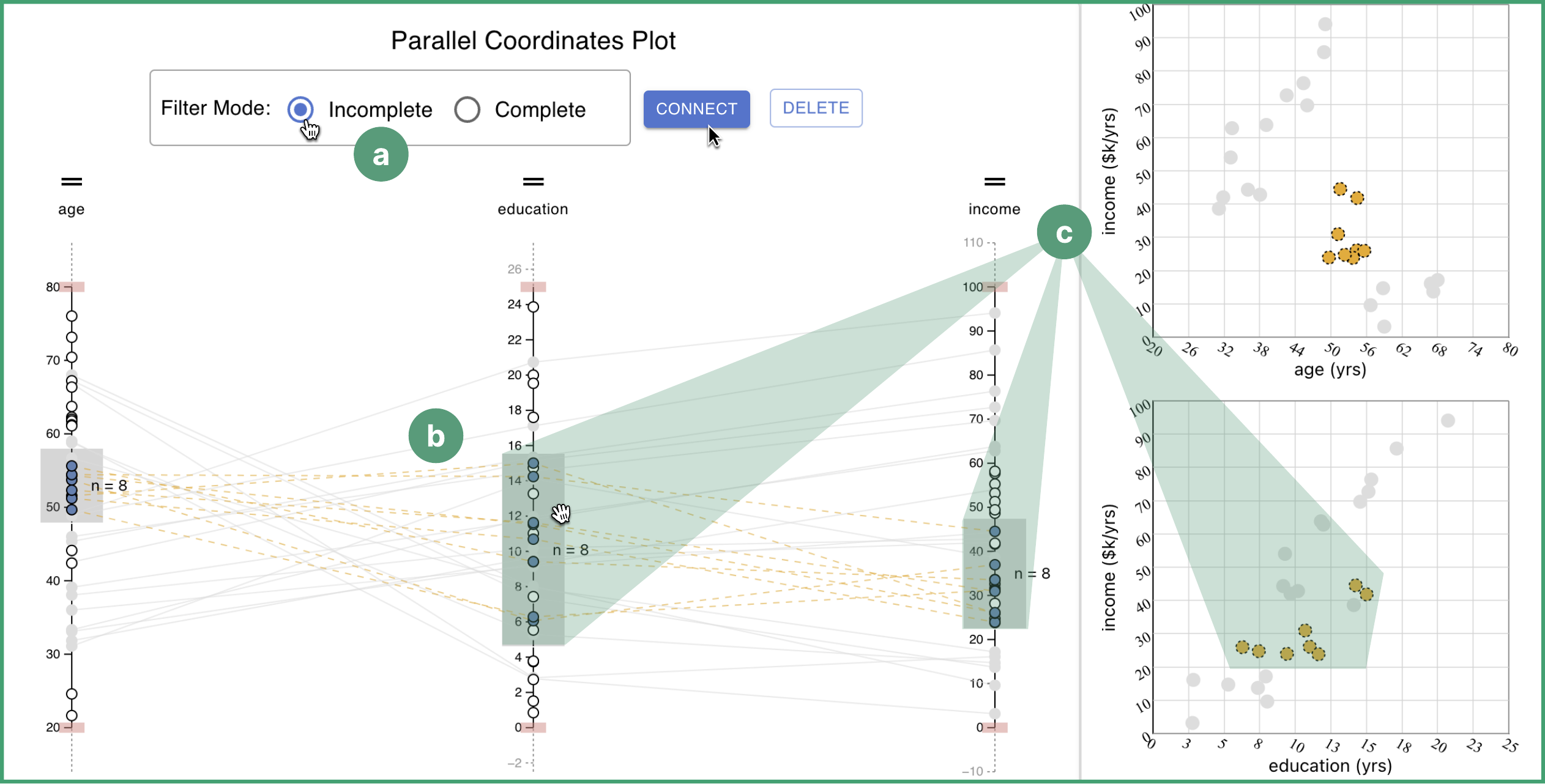}
    \caption{
    \textbf{Building relationships across multiple variables using the parallel coordinates plot and the scatterplots.} 
    (a) When analysts select the \textsc{incomplete} mode, \system{} displays only the incomplete entities (white dots) and hide the complete entities (gray dots). 
    When analysts brush regions on axes to select and connect entities, \system{} automatically identifies the maximum possible connections and previews (b) potential connections (orange dashed lines) and (c) corresponding potential entities (orange dots).
    Analysts can then click on \textsc{connect} to establish these connections and merge these entities in the underlying dataset under construction. 
    }
    \label{fig:connect}
    \Description[UI detail of the parallel coordinates plot and the scatterplots in "Incomplete" mode.]{Shows selected data points across three axes (age, education, income). Orange dashed lines and dots preview potential multivariate connections that can be finalized using a "CONNECT" button.}
\end{figure*}

To make dataset construction feasible, a key challenge is to design elicitation input methods that are both feasible for analysts to express their knowledge and sufficiently informative to derive meaningful prior distributions~\cite{mikkola2024prior}.
More specifically, analysts need interaction techniques that allow them to express assumptions about both individual variables and relationships across multiple variables in details~\cite{ohagan-2006-uncertain, gelman2020workflow}.


\system{} addresses this problem through three coordinated interactive visualizations (\autoref{fig:ui}b-f). Each view highlights a different aspect of the dataset: histograms for distributional knowledge, scatterplots for bivariate relational knowledge, and the parallel coordinates plot for multi-variable relational knowledge. Together, these views support analysts to incrementally build a dataset that reflects their assumptions (see \autoref{fig:dataset}).
Importantly, changes in one view are immediately reflected in the other views.

\subsubsection{Univariate histogram}
\label{system:visualization:distributional}

Analysts can externalize distributional assumptions of individual variables, such as plausible ranges, skew, or concentration around certain values, using univariate histograms. 
\system{} uses the histogram for specifying hypothetical samples~\cite{hullman2017imagining, kim2019bayesian} instead of allocating probabilities~\cite{oakley2010nonparametric, goldstein2014understanding} given that analysts may find it easier to think in terms of concrete values rather than probabilities~\cite{gigerenzer1995improve, hullman2017imagining, kay2016ish}. 
\revision{Each grid cell in the histogram represents a single data point within the range of its corresponding bin.
Analysts can click on a cell to add or remove data points. When adding a point, its value is randomly sampled to be within the bin’s range.}
They can also adjust the number of bins (default: $10$) or the variable’s range (default: $0$–$100$) for finer granularity.
In this way, abstract assumptions about variable distributions become tangible dataset entries, directly encoded in observable counts through frequency-based reasoning.

\subsubsection{Bivariate scatterplot}
Moreover, analysts often have expectations about how two variables relate and want a way to express those relationships~\cite{jun2024rtisane, jun2022hypothesis}. 
\revision{\system{} includes interactive scatterplots that allow analysts to brush regions to examine pairwise relationships and use the \textsc{Generate} function to add new examples within those regions. When generating examples, \system{} randomly samples values from the brushed region. Compared to interpreting line slopes in parallel coordinates—which is often difficult due to overplotting and ambiguous line crossings, scatterplots provide a direct 2D view that makes pairwise trends, clusters, and outliers easier for analysts to perceive.}
These interactions allow tacit expectations about relationships to be encoded as concrete dataset entries, while automatically linking back to the univariate histograms and forward to the parallel coordinates view.

\subsubsection{Parallel coordinates plot}

Assumptions often span multiple variables (e.g., “Older adults with high education but moderate income”), but higher-dimensional relationships are difficult to express with only univariate or bivariate views.
As a result, analysts may lose track of how their marginal or pairwise inputs interact within the dataset, producing incoherent specifications.

\system{} provides an interactive parallel coordinates plot (\autoref{fig:ui}d), where each axis corresponds to a variable and each polyline corresponds to a synthetic case (i.e., a dataset row). Analysts can brush regions on multiple axes and use the \textsc{Generate} function to add new cases that satisfy all selected ranges (see \autoref{fig:ui}e). 
\revision{New cases are constructed by randomly sampling a value within each selected range.} 
Analysts can also drag axes to reorder them. This global view extends the expressiveness of scatterplots to higher dimensions while helping analysts verify that their distributional and relational assumptions cohere at the dataset level.

\begin{figure*}[t]
    \centering
    \includegraphics[width=\linewidth]{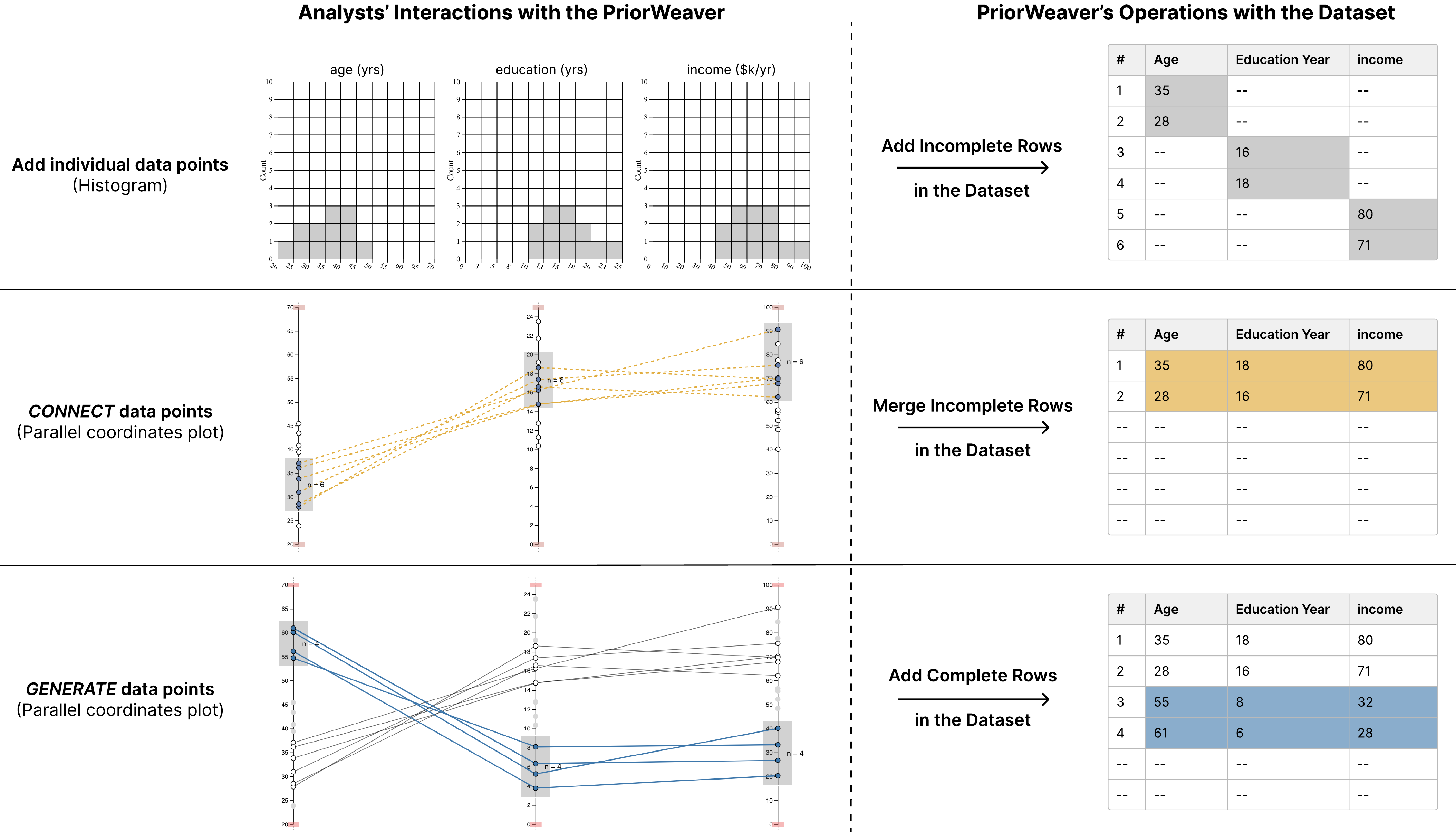}
    \caption{\textbf{Interactive dataset construction.}
    As analysts interact with the visualizations to externalize their knowledge, \system{} simultaneously constructs a dataset that represents this knowledge behind the scenes.
    The constructed dataset embodies analysts’ knowledge in two dimensions: columns record distributional assumptions about each variable, while rows link these values together, reflecting relational knowledge across variables.
    }
    \label{fig:dataset}
    \Description[Diagram illustrating interactive dataset construction steps.]{Three-row comparison: Row 1 shows adding points to a histogram creating incomplete rows in a table. Row 2 shows connecting points in a parallel coordinates plot to merge rows. Row 3 shows generating points in a parallel coordinates plot to create complete rows directly.}
\end{figure*}

\subsubsection{Coordinating multiple visualizations}
Since analysts externalize their assumptions in different views, the constructed dataset may contain both incomplete and complete entries. For example, adding values in a histogram creates incomplete rows with only one variable specified, while adding a case in the parallel coordinates plot creates a full row with values for all variables. As a result, analysts risk fragmenting their specifications into inconsistent pieces.

\system{} provides two modes---\textit{Incomplete} and \textit{Complete}---and a \textsc{Connect} function to help analysts reconcile their specifications. 
By switching modes, analysts are able to focus on either incomplete entities (rows with missing values) or complete entries (i.e., rows with full values). Entities in the unselected mode visually fade into the background.
\revision{
In the \textit{Incomplete} mode, when analysts select compatible incomplete entries in the parallel coordinates plot, \system{} generates a set of possible merges by randomly pairing incomplete entries that do not conflict with one another. It then previews possible merges by connecting paired entities with orange dashed lines and dots across the visualizations. }
This real-time visual preview enables users to validate and finalize connections (see \autoref{fig:connect}). 
%
%
Moreover, analysts can trace entries selected in one view in the other views when brushing-and-linking (\autoref{fig:ui}f).

\autoref{fig:dataset} shows how analysts leverage \system{}’s visualizations and interactions to externalize and reconcile their knowledge into a coherent, representative dataset.


\subsection{Deriving Statistical Priors}
\label{system:translation}


Analysts’ observable-level assumptions must ultimately be transformed into formal statistical priors in order to be usable in Bayesian inference. 
However, deriving priors from observable assumptions poses two difficulties. First, user-constructed datasets may contain incomplete rows with missing values, which cannot reliably contribute to parameter estimation. Second, fitting priors directly from a single constructed dataset risks overfitting and fails to capture the inherent uncertainty of priors.

\begin{figure*}[t]
    \centering
    \includegraphics[width=0.95\linewidth]{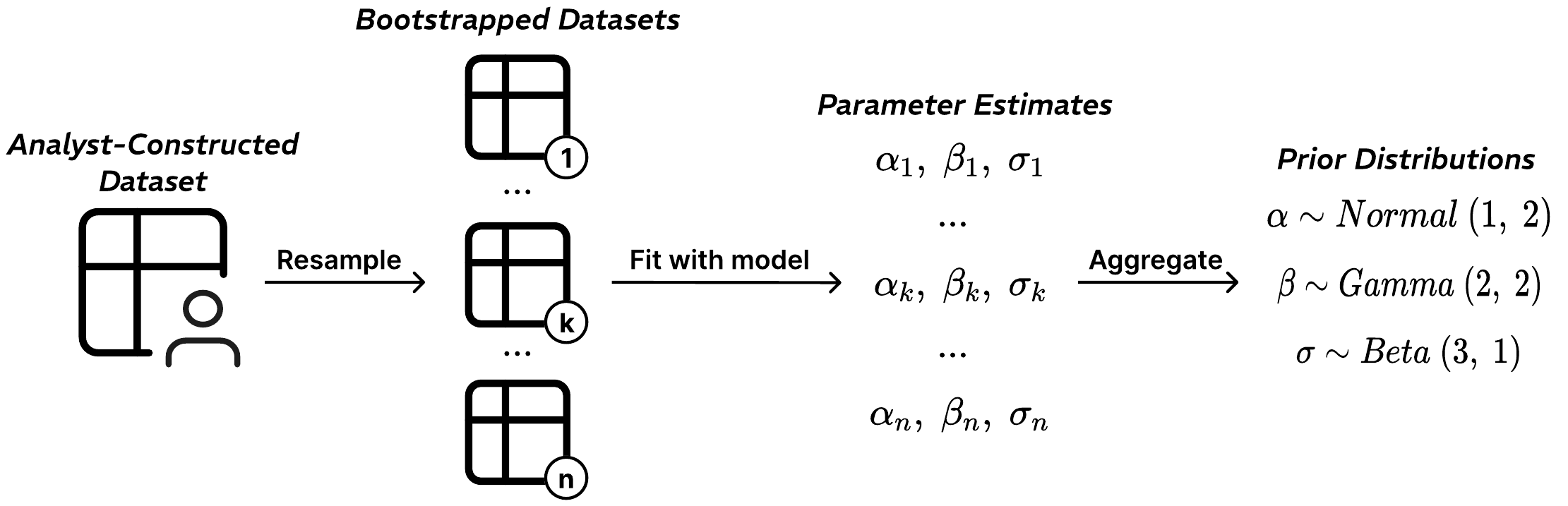}
    \caption{
    \textbf{Deriving priors from constructed dataset.}
    \system{} derives priors in three steps. (1) First, generate multiple datasets by sampling with replacement from the analyst-constructed dataset.
    (2) Next, fit the pre-specified statistical model to each bootstrapped dataset and obtain parameter estimates. 
    (3) Finally, aggregate these estimates across samples and smooth them to form continuous prior distributions.}
    \label{fig:translation}
    \Description[Flowchart of the prior derivation process.]{Three-step pipeline: (1) Resampling from the analyst-constructed dataset into $n$ bootstrapped datasets; (2) Fitting a model to each to get parameter estimates; (3) Aggregating estimates into continuous distributions like Normal, Gamma, and Beta.}
\end{figure*}

\system{} addresses both of these challenges through a three-step procedure (see \autoref{fig:translation}). 
First, \system{} filters out incomplete rows from the constructed dataset, ensuring that only fully specified cases contribute to translation.
Second, \system{} bootstraps $100$ datasets, each created by randomly sampling $50$ rows with replacement from the constructed dataset. Each bootstrapped dataset is then fit to the predefined statistical model to obtain a set of parameter estimates. This bootstrapping captures variability in the translation process and yields a more robust basis for estimation.
Finally, \system{} aggregates these parameter estimates and fits continuous probability distributions to each parameter’s possible values using Maximum Likelihood Estimation (MLE). 
\revision{
To determine the prior family, we use a predefined set of candidate distributions\footnote{The predefined set includes: Uniform, Normal, Student-t, Gamma, Beta, Skew Normal, Log Normal, Log Gamma, and Exponential.}. Using MLE, we evaluate each candidate and select the best-fitting distribution as the prior for that parameter.
}
The resulting distributions constitute the derived statistical priors used in the Bayesian analysis.

\subsection{Evaluating and Refining Derived Priors}
\label{system:refinement}

Evaluating whether derived priors reflect analysts’ knowledge is challenging because priors are defined in parameter space, but assumptions are in the observable space. To address this, \system{} employs prior predictive checks (PPCs) ~\cite{gabry2019visualization} to generate predictive distributions in the observable space for examining the consequences of prior distributions.

\begin{figure*}[t]
    \centering
    \includegraphics[width=0.97\linewidth]{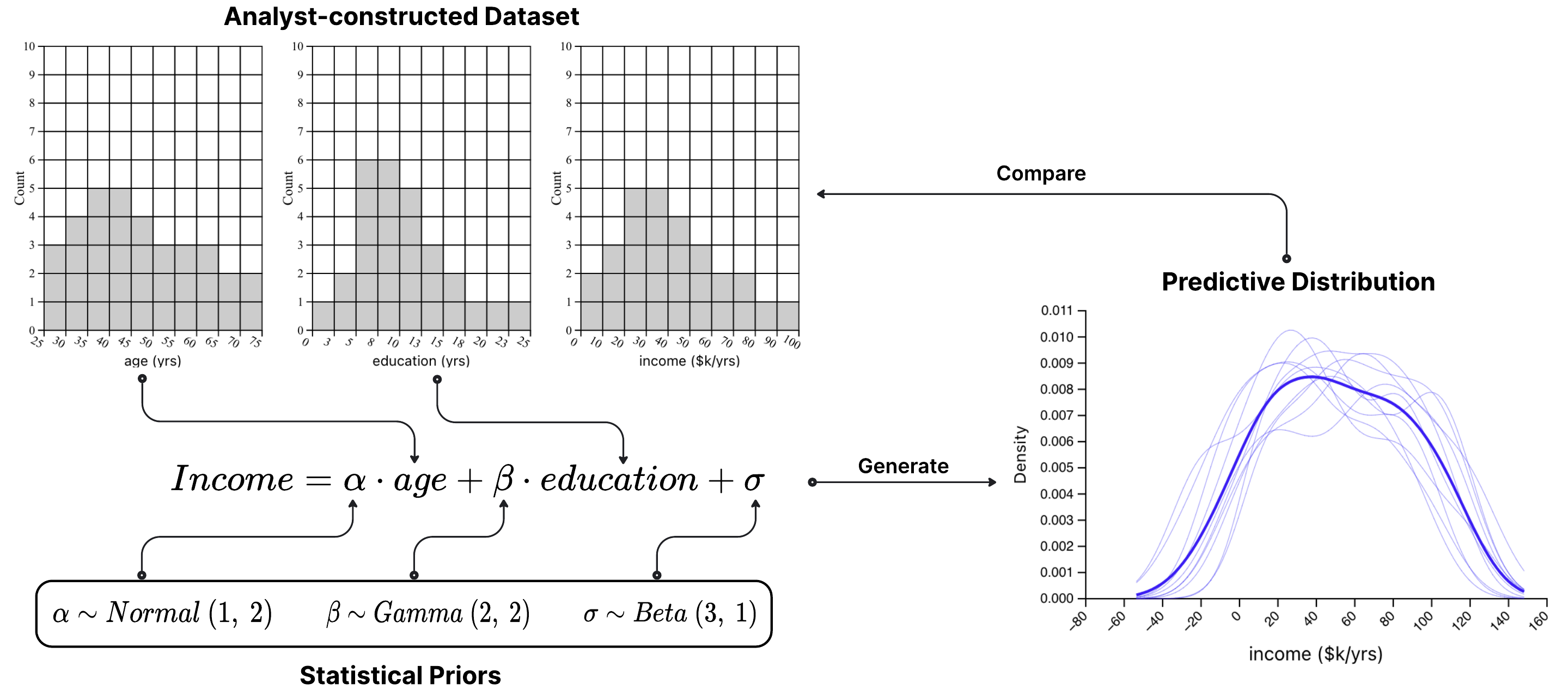}
    \caption{
    \textbf{Feedback through prior predictive checks.}
    %
    \system{} samples predictor values (e.g., age, education) from the analyst-constructed dataset (top left), draws parameter sets (e.g., $\alpha$, $\beta$, $\sigma$) from the derived priors (bottom left), and combines them to generate predictive distributions (right).
    On the rigth, each predictive distribution is shown as a faded blue line, with the average depicted as a solid blue line. 
    Analysts can detect discrepancies between these predictive distributions and the histogram of the response variable (e.g., income) (top right). 
    }
    \label{fig:feedback}
    \Description[Diagram of feedback through prior predictive checks.]{Connects the analyst-constructed dataset and derived priors to a response variable histogram. A density plot on the right compares 10 faded blue predictive distribution lines and their solid blue average against the analyst's income expectations.}
\end{figure*}

In traditional workflows, PPCs draw predictor values from an observed dataset and combine them with parameters sampled from the prior to generate predictions~\cite{gabry2019visualization}. This requires analysts to have a dataset at hand when specifying priors, which is often unrealistic and contradicts some best practices of Bayesian statistics~\cite{gelman2013philosophy}. 
%
Instead, in \system{}, predictor values are sampled directly from the analyst-constructed dataset. 
For each predictor, \system{} independently draws 100 values with replacement from its marginal distribution, as specified by the analyst’s histogram, and combines them into a simulated dataset of 100 cases. 
Then, \system{} draws 10 parameter sets from the derived priors, each of which is paired with the simulated dataset to generate a predictive distribution for the response variable. 
As a result, \system{} provides 10 predictive distributions along with their average, as illustrated in \autoref{fig:feedback}.
More importantly, analysts can directly compare the predictive distribution against the histogram of the response variable they constructed (e.g., the income histogram in \autoref{fig:feedback}) to assess alignment and identify discrepancies. 

In this way, \system{} not only offers an interpretable evaluation but also creates clear pathways for refinement. 
For example, predictive checks may reveal discrepancies, such as implausible negative incomes or extremely high average income. Analysts can return to the coordinated visualizations to adjust their assumptions by adding examples to enforce plausible ranges or re-balancing distributions.
As a result, prior elicitation becomes an iterative loop rather than a one-off specification.

\section{Usage Scenario}
\label{sec:walkthrough}

To illustrate how \system{} supports iterative prior construction, we present a usage scenario involving Jane, a social scientist investigating how age and education years influence income. 

Jane has a predefined linear model written in R ($income = \alpha \cdot age \ + \beta \cdot education\_years \ + \sigma$). 
%
She uploads the model script into \system{}, which parses the code and 
initializes blank visualizations for 
the involved variables (i.e., age, education years, income) and the parameters to be inferred (\autoref{fig:ui}a).

Jane begins by drawing on her expertise. From past research, she expects most individuals in her population to be between 25 and 55 years old, with education clustering around high school, and income broadly distributed but skewed toward lower brackets. She first records individual values for each variable in the \textbf{univariate histograms} (\autoref{fig:ui}b), adding samples that capture typical ages, education levels, and income ranges. These entries are not connected across variables. 
Next, Jane turns to the \textbf{parallel coordinates plot} (\autoref{fig:ui}d) and uses the \textbf{\textsc{connect}} (\autoref{fig:connect}) function to assemble these univariate samples into multivariate examples. This allows her to represent meaningful subgroups she has in mind, such as younger graduates with modest starting salaries or older adults with stable mid-range incomes. By switching between adding samples in the histograms and connecting entries in the parallel coordinates plot, Jane externalizes her domain knowledge as a synthetic dataset that embodies both distributions and relationships.

Once Jane has constructed this initial dataset, she clicks on the \textbf{\textsc{translate}} button. \system{} then derives prior distributions from the constructed dataset and provides a \textbf{prior predictive distribution} (\autoref{fig:ui}g) of income as feedback. To verify the behavior of derived priors, Jane compares this simulated distribution with the income histogram she specified earlier. While the two share a similar overall shape, she notices important discrepancies: the predictive distribution places too much probability on extremely high incomes. The distribution also exhibits a negative tail. These mismatches suggest that some of her earlier examples may have unintentionally overemphasized certain patterns.

Guided by this feedback, Jane revisits the visualizations. She switches to \textit{Complete} mode and adds new examples with the \textbf{\textsc{generate}} (\autoref{fig:ui}e) function. To mitigate the implausible negative tail, she generates cases of older adults with low education but still positive incomes. This reinforces her belief that income should remain above zero even with extreme cases. To counter the excessive weight on high incomes, she adds cases of middle-aged adults with high education but only moderate incomes. This reflects her belief that high education does not always yield extreme earnings.

After this new round of externalization, Jane applies \textbf{\textsc{translate}} again. This time, the predictive distribution of income better reflects her expectations: Most of the income values lie in moderate ranges. The right tail captures economic disparity without dominating the distribution. 
All income values are positive. 
Upon generating a few more potential values, Jane arrives at priors she considers both credible and well-aligned with her domain expertise.

\section{User Study}

\system{} introduces a new interactive process for prior elicitation via iterative construction of a dataset representative of analysts' domain knowledge.
\revision{
We conducted a user study to evaluate the usefulness and usability of \system{} in helping analysts externalize their domain knowledge and derive statistical priors. 
Given that previous studies have shown that setting priors using parameter-space tools is difficult for those with and without Bayesian analysis experience alike~\cite{phelan-2019-template, sarma2020prior},  
we were particularly interested in understanding the potential benefits of elicitation in the observable space. More specifically, we aimed to investigate the impact of observable-space tools on analysts without Bayesian analysis experience, including their strategies for expressing their knowledge, their abilities to evaluate and refine priors, and their interests in using Bayesian analysis in the future}.
%
%
\revision{Accordingly, we focused on the following research questions:}

\begin{itemize}
    \item \textbf{RQ1: Externalizing knowledge for priors.} 
    What strategies do analysts adopt when expressing assumptions in the observable space versus parameter space?
    How does \system{} support analysts in externalizing their implicit domain knowledge compared to a parameter-space tool? 
    
    
    \item \textbf{RQ2: Evaluating and Refining priors.} 
    How does \system{} affect how analysts evaluate and iterate on priors compared to a parameter-space tool? 

    
    \item \textbf{RQ3: Attitudes towards Bayesian analysis.} How does \system{} impact analysts' perspectives on Bayesian analysis? Specifically, are they more likely to adopt Bayesian methods using a tool like \system{}? Why or why not?
\end{itemize}

\subsection{Experimental Design}

We used a within-subjects design with interface (\system{} vs.\ baseline) and task (student performance vs.\ gym member weight) as independent variables. This design controls for participant-level differences in domain knowledge and statistical experience across interface conditions. 
\revision{
Given our interest in lowering the barrier to Bayesian analysis, we recruited analysts who have experience with statistical modeling and Frequentist analysis but are new to Bayesian methods. This population reflects the real-world users \system{} aims to support: domain experts who possess relevant knowledge but lack training in specifying priors.
}

\subsubsection{Interfaces}
\revision{To understand how elicitation spaces (parameter, observable) shape prior elicitation processes and end-user experiences, we chose to implement a parameter-space tool that incorporates prevailing prior elicitation practices as our baseline. }
As such, participants experienced two interface conditions:

\begin{itemize}
    \item \textbf{\system}: Analysts interact with the full set of features offered by our system, including expressing priors in the observable space and receiving feedback via prior predictive checks (See \autoref{fig:ui}).
    \item \textbf{Parameter space baseline}: \revision{Analysts  express their priors via the trial-roulette method~\cite{gore1987biostatistics}, which is the standard and most widely used graphical elicitation technique in the parameter space (e.g., used by \textsc{match}, \textsc{shelf} and \textsc{preliz} in Table~\ref{tab:design-space})~\cite{mikkola2024prior, falconer2022methods}. Analysts also receive feedback in the form of prior predictive checks. \autoref{fig:baseline} in the Appendix shows the baseline interface.} 
\end{itemize}





\subsubsection{Analysis tasks}
We selected two analysis tasks from prior studies on human decision-making. 
\revision{Both tasks are comparable in complexity and primarily draw on general, everyday knowledge, helping to minimize confounds from cognitive load or domain familiarity. This allows us to focus on differences in user interaction and the elicitation process.}

\begin{itemize}
    \item \textbf{Student Performance Prediction}: Analysts predict college students' exam scores based on their hours of study per week and attendance rate to class. 
    This task is adopted from previous decision-making studies \cite{dietvorst2014algo, dietvorst2018algo, weerts2019humangroundedevaluationshapalert, ghai2021explainable} and is based on a publicly available dataset on Kaggle~\cite{studentdataset}.
    \item \textbf{Gym Member Weight Prediction}: Analysts predict gym members' weight based on their height and exercise level (i.e., hours of exercise per week). This task is based on a publicly available dataset on Kaggle~\cite{gymdataset}.
\end{itemize}

\subsection{Participants}
Participants were representative of our intended user group for \system{}, analysts familiar with statistical modeling but new to Bayesian analysis. 
With approval from our institution's IRB, we recruited 17 participants
\footnote{We included the pilot participant (P1) to provide additional data. The remaining 16 participants followed the counterbalanced experimental design.}
(6 female, 11 male) 
through email and social media outreach at local universities. 
They came from diverse fields, including computer science, communication, and design. 
Participants were undergraduates, graduate students, and recent graduates currently in the workforce. 
On a five-point scale, participants self-reported moderate to substantial familiarity with the concepts of frequentist statistics ($mean=4.0, std=0.4$) and linear modeling ($mean=3.6, std=0.6$). Participants also rated their familiarity with the concepts behind Bayesian analysis (e.g., Bayes' theorem) as slight to moderate ($mean=2.7, std=0.2$), but, importantly, none of the participants had performed a Bayesian analysis before.
%
The study lasted approximately 80 minutes, and participants received a \$25 USD gift card for their time.

\begin{table}[t]
\centering  

\caption{\textbf{Participant demographics and background.}}  

\begin{tabular}{lclllc}
\toprule
\textbf{PID} &
  \textbf{Age} &
  \textbf{Gender} &
  \textbf{Field} &
  \textbf{Study/Job} &
  \\ \hline
P1 &
  \cellcolor[HTML]{FFFFFF}{\color[HTML]{434343} 27} &
  \cellcolor[HTML]{FFFFFF}{\color[HTML]{434343} Male} &
  \cellcolor[HTML]{FFFFFF}{\color[HTML]{434343} HCI} &
  \cellcolor[HTML]{FFFFFF}{\color[HTML]{434343} PhD student} &
  \\ 
P2 &
  \cellcolor[HTML]{FFFFFF}{\color[HTML]{434343} 22} &
  \cellcolor[HTML]{FFFFFF}{\color[HTML]{434343} Male} &
  \cellcolor[HTML]{FFFFFF}{\color[HTML]{434343} Computer Science} &
  \cellcolor[HTML]{FFFFFF}{\color[HTML]{434343} Master's student} &
  \\ 
P3 &
  \cellcolor[HTML]{FFFFFF}{\color[HTML]{434343} 23} &
  \cellcolor[HTML]{FFFFFF}{\color[HTML]{434343} Male} &
  \cellcolor[HTML]{FFFFFF}{\color[HTML]{434343} Computer Science} &
  \cellcolor[HTML]{FFFFFF}{\color[HTML]{434343} PhD student} &
  \\ 
P4 &
  \cellcolor[HTML]{FFFFFF}{\color[HTML]{434343} 23} &
  \cellcolor[HTML]{FFFFFF}{\color[HTML]{434343} Male} &
  \cellcolor[HTML]{FFFFFF}{\color[HTML]{434343} Computer Science} &
  \cellcolor[HTML]{FFFFFF}{\color[HTML]{434343} PhD student} &
  \\ 
P5 &
  \cellcolor[HTML]{FFFFFF}{\color[HTML]{434343} 23} &
  \cellcolor[HTML]{FFFFFF}{\color[HTML]{434343} Female} &
  \cellcolor[HTML]{FFFFFF}{\color[HTML]{434343} HCI} &
  \cellcolor[HTML]{FFFFFF}{\color[HTML]{434343} Master's student} &
  \\ 
P6 &
  \cellcolor[HTML]{FFFFFF}{\color[HTML]{434343} 35} &
  \cellcolor[HTML]{FFFFFF}{\color[HTML]{434343} Male} &
  \cellcolor[HTML]{FFFFFF}{\color[HTML]{434343} HCI} &
  \cellcolor[HTML]{FFFFFF}{\color[HTML]{434343} Postdoc} &
  \\ 
P7 &
  \cellcolor[HTML]{FFFFFF}{\color[HTML]{434343} 23} &
  \cellcolor[HTML]{FFFFFF}{\color[HTML]{434343} Male} &
  \cellcolor[HTML]{FFFFFF}{\color[HTML]{434343} Computer Science} &
  \cellcolor[HTML]{FFFFFF}{\color[HTML]{434343} Master's student} &
  \\ 
P8 &
  \cellcolor[HTML]{FFFFFF}{\color[HTML]{434343} 21} &
  \cellcolor[HTML]{FFFFFF}{\color[HTML]{434343} Male} &
  \cellcolor[HTML]{FFFFFF}{\color[HTML]{434343} Data Science} &
  \cellcolor[HTML]{FFFFFF}{\color[HTML]{434343} Undergraduate} &
  \\ 
P9 &
  \cellcolor[HTML]{FFFFFF}{\color[HTML]{434343} 30} &
  \cellcolor[HTML]{FFFFFF}{\color[HTML]{434343} Female} &
  \cellcolor[HTML]{FFFFFF}{\color[HTML]{434343} HCI} &
  \cellcolor[HTML]{FFFFFF}{\color[HTML]{434343} Researcher} &
  \\ 
P10 &
  \cellcolor[HTML]{FFFFFF}{\color[HTML]{434343} 22} &
  \cellcolor[HTML]{FFFFFF}{\color[HTML]{434343} Female} &
  \cellcolor[HTML]{FFFFFF}{\color[HTML]{434343} HCI} &
  \cellcolor[HTML]{FFFFFF}{\color[HTML]{434343} Master's student} &
  \\ 
P11 &
  \cellcolor[HTML]{FFFFFF}{\color[HTML]{434343} 30} &
  \cellcolor[HTML]{FFFFFF}{\color[HTML]{434343} Male} &
  \cellcolor[HTML]{FFFFFF}{\color[HTML]{434343} Communications} &
  \cellcolor[HTML]{FFFFFF}{\color[HTML]{434343} Master's student} &
  \\ 
P12 &
  \cellcolor[HTML]{FFFFFF}{\color[HTML]{434343} 23} &
  \cellcolor[HTML]{FFFFFF}{\color[HTML]{434343} Male} &
  \cellcolor[HTML]{FFFFFF}{\color[HTML]{434343} Computer Science} &
  \cellcolor[HTML]{FFFFFF}{\color[HTML]{434343} PhD student} &
  \\ 
P13 &
  \cellcolor[HTML]{FFFFFF}{\color[HTML]{434343} 18} &
  \cellcolor[HTML]{FFFFFF}{\color[HTML]{434343} Male} &
  \cellcolor[HTML]{FFFFFF}{\color[HTML]{434343} Data Science} &
  \cellcolor[HTML]{FFFFFF}{\color[HTML]{434343} Undergraduate} &
  \\ 
P14 &
  \cellcolor[HTML]{FFFFFF}{\color[HTML]{434343} 27} &
  \cellcolor[HTML]{FFFFFF}{\color[HTML]{434343} Female} &
  \cellcolor[HTML]{FFFFFF}{\color[HTML]{434343} Design} &
  \cellcolor[HTML]{FFFFFF}{\color[HTML]{434343} Designer} &
  \\ 
P15 &
  \cellcolor[HTML]{FFFFFF}{\color[HTML]{434343} 22} &
  \cellcolor[HTML]{FFFFFF}{\color[HTML]{434343} Female} &
  \cellcolor[HTML]{FFFFFF}{\color[HTML]{434343} Data Science} &
  \cellcolor[HTML]{FFFFFF}{\color[HTML]{434343} Undergraduate} &
  \\ 
P16 &
  \cellcolor[HTML]{FFFFFF}{\color[HTML]{434343} 27} &
  \cellcolor[HTML]{FFFFFF}{\color[HTML]{434343} Female} &
  \cellcolor[HTML]{FFFFFF}{\color[HTML]{434343} HCI} &
  \cellcolor[HTML]{FFFFFF}{\color[HTML]{434343} PhD student} &
  \\ 
P17 &
  \cellcolor[HTML]{FFFFFF}{\color[HTML]{434343} 28} &
  \cellcolor[HTML]{FFFFFF}{\color[HTML]{434343} Male} &
  \cellcolor[HTML]{FFFFFF}{\color[HTML]{434343} Robotics} &
  \cellcolor[HTML]{FFFFFF}{\color[HTML]{434343} Master's student} &
  \\
\bottomrule
\end{tabular}
\label{tab:participant}
\vspace{-\baselineskip}
\end{table}



\subsection{Procedure}

After obtaining informed consent, a researcher introduced the study’s objectives and overall procedure. Participants then completed a pre-task survey assessing their background in statistics and gathering demographic data.

Participants completed two analysis tasks, each using a different interface. Task order and interface assignment were counterbalanced and randomly assigned to participants.
For each task, participants followed the same five-step process:

\begin{itemize}
    \item[1.]  \textbf{Tutorial \& Practice}. Participants first watched a tutorial video introducing the interface (\system{} or baseline). 
        Participants then freely explored the interface until they felt familiar with it.
        The tutorials for both interfaces used the same analysis example task, which was to 
        predict income based on age and years of education. 
        We chose this task for the tutorials since it is distinct enough from the experimental tasks and has been widely used in previous
        decision-making research \cite{zhang2020effect, hase2020evaluating, ribeiro2018anchors, ghai2021explainable}.
    \item[2.] \textbf{Task Introduction}. A researcher introduced the
    analysis task (student performance or gym member weight).
    \item[3.] \textbf{Prior Elicitation}. 
    \revision{After confirming their understanding of the task, participants iteratively specified their priors using the assigned interface. In each iteration, they expressed their assumptions, generated priors, and examined the resulting predictive distributions. They repeated this process until 
    they judged the results to adequately reflect their beliefs.}
    \item[4.] \textbf{Post-Task Survey}. 
    Participants completed a questionnaire related to their experience with the current condition.
    \item[5.] \textbf{Post-Task Interview.} 
    A researcher asked participants open-ended questions about their experience with the system and the analysis task.
\end{itemize}

After completing all five steps for the first task, participants proceeded to
the second task. After both tasks, 
a researcher engaged participants in a final semi-structured exit 
interview about additional reflections and feedback on their
experiences. 
All study materials are included as supplemental material.

\subsection{Measurements and Analysis}
\label{evaluation:measurements}

We analyzed data from system interaction logs, surveys, and semi-structured interview transcriptions. 
\secondRevision{In addition, we examined the priors and associated predictive distributions produced at each elicitation iteration, with a particular focus on differences between the initial and final rounds.}

Quantitative metrics are based on participants’ survey responses and interaction logs. The data violated the parametric assumption of normality, so we conducted Wilcoxon signed-rank tests for paired comparisons between interface conditions. We applied the Benjamini–Hochberg correction to account for multiple comparisons.

We conducted a thematic analysis of the open-ended survey responses and semi-structured interview transcriptions. 
Our pre-defined interview questions guided the coding process. 
Two authors independently coded the transcripts, developed a shared codebook, and resolved discrepancies through discussion.

\section{Findings}
\label{sec:findings}


\begin{figure*}[t!]
    \centering
    \includegraphics[width=\linewidth]{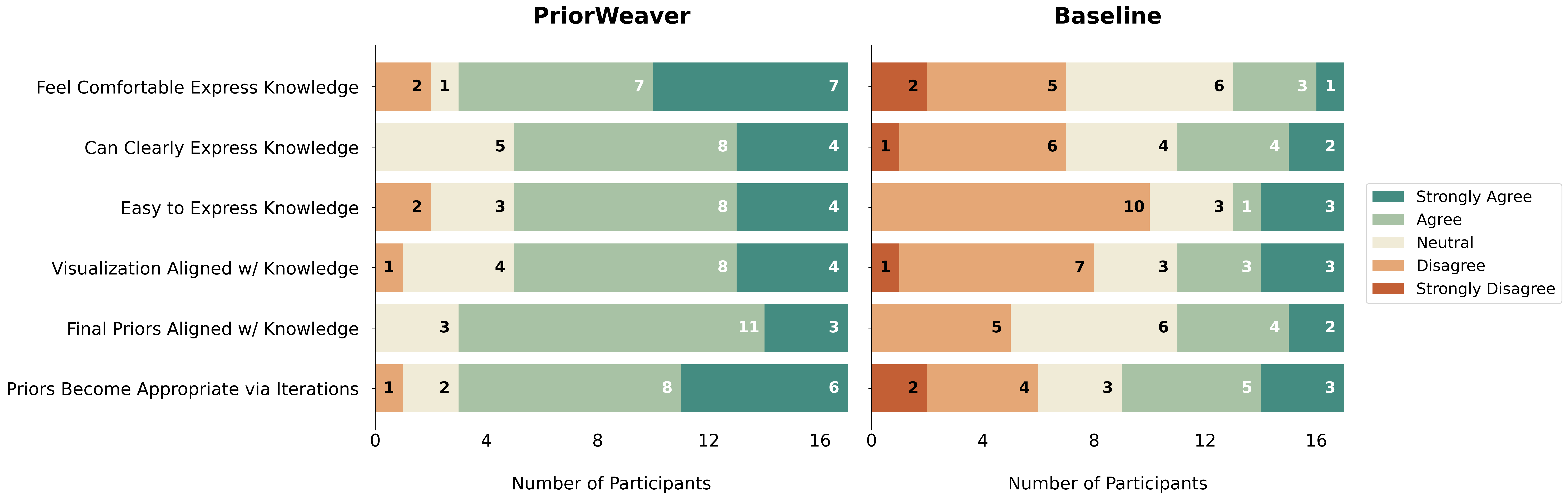}
    \caption{
    \textbf{Participants' survey responses comparing \system{} and the parameter-space baseline across six dimensions.}
    Results indicate that \system{} provided stronger support for analysts to articulate their domain knowledge, obtain priors aligned with their expectations, and refine those priors effectively through iteration.}
    \label{fig:findings:survey-results}
    \Description[Likert scale survey results comparing \system{} to the baseline.]{Diverging bar charts for six metrics. \system{} shows significantly higher "Strongly Agree" and "Agree" counts for comfort, clarity, ease of expression, and alignment compared to the baseline, which has more "Disagree" ratings.}
\end{figure*}

\subsection{RQ1: Strategies for Domain Knowledge Externalization}

\subsubsection{There are three strategies for externalizing domain knowledge in the observable space using \system{}.}

Participants often began with a single strategy, commonly distribution-driven or example-driven, and then flexibly switched across strategies, moving back and forth as needed to articulate their knowledge.

\begin{itemize}
    \item \textbf{Distribution-driven strategy: Matching marginal shapes.}
    When participants had clear expectations about how individual variables should behave, they focused on shaping the marginal distributions of those variables. For instance, they adjusted data points in histograms until a distribution's center, spread, or skew, aligned with their beliefs.
    \item \textbf{Association-driven strategy: Matching pairwise relationships.}
    When participants thought about how variables relate to one another, they concentrated on expressing pairwise associations. For example, they used the parallel coordinates plot and scatterplots to represent and verify their assumptions about the slope and form of relationships (e.g., positive, negative, linear, or nonlinear) between variables.
    \item \textbf{Example-driven strategy: Matching multivariate patterns.}
    When participants anchored their reasoning in specific, concrete real-world examples, they specified data points across variables to represent an individual example. For instance, they drew out a single plausible case. They used the parallel coordinates plot to encode multivariate patterns that captured how variables combine to form one entity observable in the real-world.
\end{itemize}


The vast majority of participants (14/17) began their externalization with the distribution-driven strategy, focusing first on histograms. As P7 explained, \userQuote{it felt natural to start with the histogram and think about each variable in isolation.}
Similarly, P1 noted that \userQuote{I drew the histogram first, so that I could be more purposeful when connecting different variables.}
From there, some participants (P1, P12, P13) moved to the relationship-driven strategy, emphasizing that they \userQuote{felt confidence in pairwise relationships} (P13) and that \userQuote{pairwise relationships were very direct} (P12). 
Others (P3, P5–8, P10–17) transitioned to the example-driven strategy. As P7 described, \userQuote{when the [data points] were established from histograms, I would then think about the different groups of real-world examples that I had in mind, such as low X + high Y = high Z.}

A few participants (P2, P4, P9) started with the example-driven strategy, creating multivariate examples in the parallel coordinates plot. 
P9 described how relational knowledge guided his process: \userQuote{Rather than distributions, some relational knowledge (i.e., examples) came to mind right away when I saw the task. So I expressed them in the parallel coordinates plot, then worked backward to the histograms to see if the distributions made sense.}
Similarly, P4 explained his reasoning in detail: \userQuote{To my knowledge, students who study more hours and have a higher attendance rate usually get better final exam scores. So I used \textsc{generate} to create these examples first. And I have an approximate mean value and shape of distribution for study hours in mind, so I built the distribution next.}

\subsubsection{With the baseline tool, participants struggle to follow a clear strategy for externalizing their knowledge.}
The majority of participants (10/17) relied on guessing through trial and error when using the baseline tool. 
As P5 described, \userQuote{[I] would randomly guess a mean value and a distribution shape, then tweaked them until the predictive distribution looked appropriate.}
The other seven participants vocalized beliefs in observable space terms first then manually translated these into the parameter space. 
P7, P13, P14, and P17 performed \userQuote{rough mental conversion} (P14).
P3, P11, and P12 completed \userQuote{manual calculation[s] on paper} (P3) to estimate plausible mean values and ranges for parameters. 
P3 explained, \userQuote{I set up an equation that if someone has maximum study hours and attend[s] every class [they] should have a 100 exam score. Based on that, I estimated that both parameters’ mean should be around 0.5.}
These approaches demonstrate that, without \system{}, participants were required to bear the substantial cognitive burden of transforming their beliefs into statistical parameters.

Echoing these observations, participants reported that \system{}'s visualizations were well aligned with their knowledge ($Z$ = -2.280, $p$ < 0.01), reducing the cognitive effort required to express their knowledge and allowing them to apply purposeful strategies.

\subsection{RQ1: General Support for Knowledge Externalization}
\label{findings:expression}


\subsubsection{The parameter space lacks support for translation of domain knowledge.}

In the baseline condition, most participants (15/17) struggled to express their knowledge about parameters. 
They described the parameter space as \userQuote{too abstract} (P17) and admitted being \userQuote{confused about the meaning of parameter distributions} (P3). 
As P4 explained, they \userQuote{have knowledge about the variables, but it is hard to correctly and clearly transform the knowledge about the variables into knowledge about the parameters.} 
Similarly, P11 remarked that \userQuote{there is a gap between what I knew and what I need to express.}
%
Beyond understanding individual parameters, participants (P2, P3, P7, P9, P17) also found it difficult to reason about them jointly. 
Even with a background in machine learning, P2 explained, \userQuote{it is difficult to set parameters individually and consider the combined effects at the same time. This often lead[s] to unexpected results.}

Interestingly, when asked how to cope with these challenges, seven participants (P3, P7, P11-14, P17) reported that they would reason about the variables rather than the parameters themselves. 
For example, some described that they would \userQuote{think about the correlation between a variable and the outcome} (P12) 
or \userQuote{do a rough mental estimate of how each variable would affect the outcome} (P17). 
Similarly, P3 explained that he would \userQuote{set up an equation and fill in variable values to calculate the possible values of parameters.}
In other words, participants were imagining and simulating manipulations in the observable space. 

\subsubsection{The observable space makes knowledge externalization more direct.}
With \system{}, all participants appreciated being able to externalize their knowledge in the observable space directly. 
They explained that this approach \userQuote{aligns better with [their] natural thinking process} (P11) and \userQuote{abstracts away the need to deal with parameters} (P7). 
Several described the process as \userQuote{intuitive} (P3, P4, P6, P10, P12, P17) because they can \userQuote{input real values from examples I encounter in daily life} (P10) and \userQuote{reason with frequency format} (P6).
For instance, P3, who shared that he frequently goes to the gym, reasoned about the weight prediction task by recalling \userQuote{several figures and samples in my mind, of people who I regularly see in the gym.}
In this way, \system{} allowed participants to focus on surfacing and expressing their domain knowledge, rather than figuring out how to represent it as parameters. 

Indeed, as shown in \autoref{fig:findings:survey-results}, participants gave significantly higher ratings to \system{} than to the baseline in terms of 
comfortable of expression ($Z$ = -2.866, $p$ < 0.05), 
clarity of expression ($Z$ = -2.684, $p$ < 0.01) and ease of expression ($Z$ = 3.022, $p$ < 0.01).

\subsubsection{\system{} provides better support for expressing knowledge about variable relationships.}

Ten participants (P2-5, P7-9, P11, P12, P16) valued how \system{} enabled them to see and build relationships between variables more \userQuote{clearly} (P7) and \userQuote{easily} (P8). 
Participants found \system{}'s parallel coordinates plot particularly helpful. 
P2 elaborated on how \system{} supported them: 
\userQuote{It’s hard to express the final outcome based on each variable separately. You have to combine them together and express that kind of knowledge. I think the parallel coordinates plot, which allows me to \textsc{connect} or \textsc{generate} the data points, helps me effectively express the combined effects of multiple variables.}
P5 also noted how the parallel coordinates plot \userQuote{helped capture knowledge that might otherwise be overlooked if expressed only through histograms}. 
Furthermore, three participants (P7, P9, P13) highlighted the usefulness of the bivariate scatterplot. P7 found that it was \userQuote{a great breakdown of what the parallel coordinates plot is trying to show} and allowed them to \userQuote{focus on two variables at a time and validate: Does this relationship make sense?}

At the same time, the requirement to express variable relationships was challenging for some participants (P11, P12, P13, P15).
For instance, P13 explained that they had beliefs about only some subset of the variables, so \userQuote{it was hard to find clear relationships between all variables.} 
Related, four participants (P4, P7, P15, P17) remarked that the parallel coordinates plot and its functions made \system{} more tedious for externalizing knowledge and more challenging to learn initially compared to the baseline tool.


\begin{figure*}[t!]
    \centering
    \includegraphics[width=\linewidth]{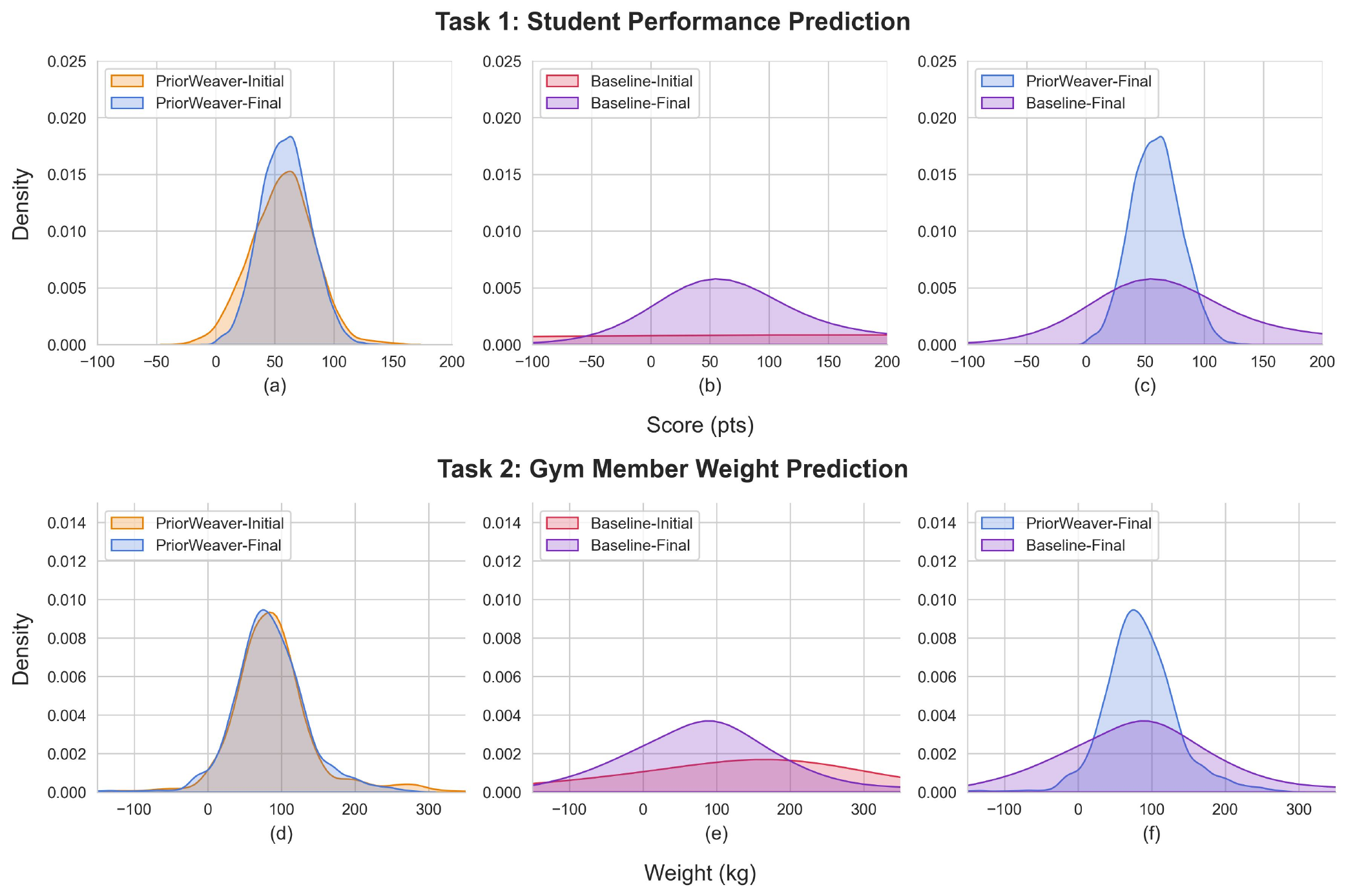}
    \caption{\textbf{Participants specified priors that produced more reasonable predictive distributions when using \system{}.} 
    \revision{
    \textit{Initial} and \textit{Final} refer to the first and last elicitation iterations, respectively.  
    Predictive distributions are aggregated across participants. 
    Figures (a) and (d) show initial versus final predictive distributions in \system{}. 
    Figures (b) and (e) show initial versus final predictive distributions in the baseline. 
    Figures (c) and (f) compare final predictive distributions between \system{} and the baseline.
    Across both tasks, participants using \system{} produced predictive distributions that aligned more closely with their expectations from the first iteration and ultimately arrived at more reasonable final predictive distributions than those obtained with the baseline.
    }}
    \label{fig:ppd-results}
    \Description[Density plots comparing initial and final predictive distributions.]{Six sub-plots (a–f) for two tasks. \system{}'s initial and final distributions are tightly aligned and bell-shaped, while the baseline's initial distributions are much flatter and broader, showing more drastic shifts after iteration.}
\end{figure*}

\subsection{RQ2: Evaluating and Refining Priors}




\subsubsection{Participants use similar criteria for assessing prior quality in both tools.}
%
Participants  (P2, P3, P4, P7, P9, P11-14, P17) paid attention to the ranges and tails of the predictive distributions. They checked for any unrealistic extreme values. In the exam score prediction task, five participants (P3, P7, P9, P13, P17) mentioned \userQuote{if there is notable density of the distribution exceed 100, then [they] would refine another round} (P7).
%
When visually assessing the prior predictive distribution, five participants (P4, P5, P6, P8, P14) looked to see that \userQuote{the majority of [the] density should fall between (or around) [certain] interval[s] that approximately matches with what [they] expressed} (P4). Three participants (P7, P12, P17) focused more on the \userQuote{overall shape of the distribution} (P17).


\subsubsection{In the baseline, refinement is frustrating and unpredictable.}
%
The majority of participants (P1, P3-6, P8, P10, P14, P16, P17) struggled to identify what to adjust and how even after recognizing that the predictive distributions were not aligned with their expectations. Participants ultimately resorted to trial-and-error. 
For example, P3 explained, \userQuote{The results still show scores over 100. I know I should have eliminated that part, but I didn’t know how to do it.}

Six participants (P2, P7, P9, P11, P12, P13) tried 
\userQuote{narrowing parameter ranges to reduce unrealistic values} (P13) or \userQuote{lowering peak parameter values to lower the effect} (P7).
However, these adjustments often failed to produce the expected results. 
As P2 explained, \userQuote{even small changes in parameter values could lead to large and unintended shifts in the predictive distribution.}

\subsubsection{In \system{}, iteration is more purposeful.}
When participants noticed undesired tails or ranges in the predictive distribution, they (P1, P4, P5, P6, P9, P10, P15) responded by adding specific data points to capture extreme scenarios or by removing unrealistic cases. 
A couple of participants (P10, P14) also chose to add data points that counterbalanced the results. 
P4's reflection on their refinement process with \system{} succinctly summarized its main benefit for iteration: \userQuote{[I] could clearly locate where anomalies were and have a concrete direction for refinement.}

For instance, to shift distributions with undesired shapes, six participants (P2, P6, P11, P12, P13, P17) modified specific intervals.
%
P11 described, \userQuote{The predicted average score was too low, so I linked some hardworking but low-attendance students with high scores, which might increase the overall exam scores.}
P2 had a similar approach: \userQuote{The average grade should be between 70 and 80, but the distribution was uniform. So I added more points around that specific grade interval.}
%
In addition, participants (P2, P4, P7, P13) refined the strength of relationships. 
As P7 put it, \userQuote{The number of points can represent the ‘weight’ of relationships in the model.}
Using the \textsc{generate} function, P7 emphasized how it allowed them to \userQuote{quickly establish new relationships or emphasize certain relationships on the fly.}


When comparing the baseline and \system{}, participants reported that the final elicited priors in \system{} were more aligned with their knowledge ($Z$ = -2.397, $p$ < 0.05). 
%
\secondRevision{We examined the priors and corresponding predictive distributions from the initial and final elicitation iterations and observed that, by the final iteration, participants using \system{} produced more appropriate prior predictive distributions than those using the baseline.
For instance, most predicted scores fell between 20 and 100 points, and most predicted weights fell between 40 and 150 kilograms.
In addition, when using \system{}, participants’ initial priors produced predictive distributions that were both more appropriate and more closely aligned with the final predictive distributions derived from their final priors.}
\autoref{fig:ppd-results} illustrates this. Put simply, \system{} enabled analysts to specify reasonable priors right away and refine them meaningfully, resulting in priors that were better-aligned with their knowledge overall.

\subsection{RQ3: Attitudes Towards Bayesian Analysis}
\label{findings:experience}


Participants reported that \system{} made the complex task of Bayesian prior elicitation more approachable for them (P2-8, P10, P14).
%
P14 summarized that the tool \userQuote{breaks down Bayesian analysis into bite-sized chunks, making the process far less overwhelming than learning from textbooks.}
Similarly, P7 appreciated that the system allowed him to \userQuote{quickly get started exploring relationships without heavy technical setup,} highlighting its support for exploratory and iterative thinking.
%
Participants found
\system{} significantly more helpful for externalizing their knowledge compared to
the baseline ($Z$ = -3.169, $p$ < 0.01). 
They also reported higher confidence in using \system{} ($Z$ = -2.939, $p$ < 0.01) and found it less cumbersome to use ($Z$ = -2.359, $p$ < 0.05).



Participants (P2, P3, P11, P14) also expressed that \system{} transformed Bayesian analysis
from a theoretical concept into a practical tool for real-world use.
%
%
P3 emphasized, \userQuote{If I don't have these
tools, Bayesian analysis would only exist in textbooks for me.} 
Additionally, P11 
shared, \userQuote{This process really gave me an introduction to
Bayesian analysis. It made me realize it's something I can actually apply in my future research.} Participants were significantly more
inclined to apply Bayesian methods in future projects if they had access to
\system{} than the baseline ($Z$ = -2.676, $p$ < 0.01).
%

\subsection{Opportunities to Improve \system{}}


Participants gave feedback on ways to improve \system{}.
%
%
First, a couple participants (P6, P11) were unfamiliar with parallel coordinates plots and asked for clearer onboarding materials, simpler explanations of system features (e.g., the distinction
between complete and incomplete modes), and lightweight tutorial videos to
support familiarization.
%
Second, participants (P1, P10, P15) wanted more insight into and control over the entire
process. Participants asked for more transparency into the process of turning the visual inputs into statistical priors (P1), finer histogram bins (P10), and recommended parameter ranges to support refinement (P15).
%
%
%
Third, participants hoped for more flexible ways to express their knowledge, including incorporation of categorical variables (P7), nonlinear variable relationships (P16), and multiple input modalities (e.g., rule-based logic, natural language) (P9).



\subsection{Key Takeaways}
\system{} scaffolded the elicitation process by guiding participants through knowledge externalization and prior iteration. For both, \system{} gave participants a greater sense of control in formulating goals, working towards them, and assessing their progress. During knowledge externalization, \system{}'s use of the observable space allowed participants to accurately express their knowledge, while employing multiple possible strategies. As a result, participants reported feeling more comfortable and clearer about what and how to express, leading to initial priors that were more reasonable and better aligned with their expectations.
While iterating on priors, participants leveraged prior predictive checks to evaluate priors in both tools. Yet, \system{} made the feedback more actionable and facilitated more goal-oriented refinements. Without \system{}, participants relied on trial-and-error changes. Overall, \system{} lowered the barriers to prior elicitation and Bayesian analysis for participants.

\section{Discussion} \label{discussion}

This paper develops and evaluates \system{}, an interactive system for prior
elicitation via iterative dataset construction. \system{} scaffolds elicitation
by guiding participants to externalize knowledge in the observable space and
structures prior iteration around predictive checks. In a lab study, we found
that \system{} helped participants express their knowledge more systematically
and directly. It also helped participants refine priors more effectively
than with trial-and-error approaches common with the parameter-space baseline.
Our design process and the evaluation results give insight into the design of abstractions for eliciting beliefs and the role of constructed datasets as knowledge representations. 

\subsection{Balancing Usability and Control in Belief Elicitation} \label{discussion:tradeoff}

With \system{}, participants in the user study could easily specify distributions and relationships in the observable space. The underlying parameterization of the model was abstracted away. As a result, they could not directly change exact coefficients or distributional forms.
In contrast, the parameter-space baseline system gave participants fine-grained control, but most participants struggled to
wield this control effectively due to a lack of Bayesian analysis expertise. Many
reported that adjusting parameters felt like \textit{``guessing''} (P5), and
changes often led to unintended consequences in the predictive check results. 
Based on these results, we conclude that trading off control for ease of expression is the right design decision for Bayesian novices. 
%

Generalizing beyond Bayesian analysis, our evaluation suggests that interfaces should minimize the need for users to translate their beliefs and provide feedback in forms that match the input to make it more actionable.
Without this kind of support, users have the burden of manually translating their implicit beliefs into terms appropriate for the task. This process is not only likely to be error-prone but also lead some to resort to trial-and-error.

\subsection{Constructed Datasets as Knowledge Representations}

\system{}'s design intends for users to engage with their implicit beliefs (and gaps therein) through interacting with the coordinated visualizations. In the user study, participants adopted different strategies based on what they knew about the domain. 
For example, some started with distributions of variables. Others thought through concrete examples across all variables. In other words, \system{} scaffolded how analysts recollected their domain knowledge. Through iteration, the constructed datasets came to capture a data generating process for the domain. 

Furthermore, the dataset that analysts create through the interactive visualizations serves as an intermediate representation between analysts’ abstract domain knowledge and their statistical priors. 
\revision{Viewed in this light, the constructed dataset may be useful for other stages of the data lifecycle, including model iteration.
For example, following the Bayesian workflow~\cite{gelman2020workflow}, analysts may revise their model after eliciting priors in \system{}. In such cases, analysts could reuse the constructed dataset underlying \system{} to explore alternative model specifications and evaluate model behavior before examining empirical data, mitigating concerns around ``double dipping,'' which can inflate false discovery rates. 
In this way, the constructed dataset could serve as a shared representation~\cite{heer2019agency} between analysts and systems, increasing interpretability, transparency, and reproducibility.}

\subsection{Benefits of Concretizing Implicit Assumptions}
\system{} grounded analysts’ thinking in concrete, familiar representations. 
Specifically, its interactive visualization
interface provided a higher level of abstraction than existing parameter-space
prior elicitation tools. 
The result was that prior elicitation felt less like performing abstract statistical modeling and more like expressing analysts' knowledge. 
In the study, participants felt greater confidence and expressed less confusion
when using \system{}, even without additional guidance or previous Bayesian analysis
experience~\cite{phelan-2019-template}.
These observations align with prior research showing that interfaces which reflect users’
mental models and everyday reasoning can foster greater confidence and
engagement, particularly in complex or technical domains
\cite{kulesza2015principles, hullman2019authors}. 

%



\subsection{Lowering the Barrier to Bayesian Analysis}
Beyond supporting prior elicitation process, \system{} helped shift
participants' broader attitudes toward Bayesian analysis. Participants, all of whom
initially had little or no experience with Bayesian methods, expressed greater
willingness and confidence to apply Bayesian approaches in the future after
using the system. They emphasized how the visual and interactive nature of the
interface in the observable space made prior elicitation feel approachable and actionable. 
This is in stark contrast to the baseline system focused on the parameter-space that participants found to be 
abstract and theoretical. 
Several participants noted that without tools like \system{}, Bayesian methods
would have remained confined to textbooks and theoretical discussions.
\revision{
In this light, \system{} could also function as a pedagogical instrument, helping facilitators teach domain experts not only how to set priors but also how to reason about key tradeoffs, such as the level of informativeness.
}

\section{Limitations and Future Work} \label{sec:limitations}

Our longer term goal is to make Bayesian analysis more approachable for analysts
of various backgrounds in real-world scenarios. Towards this goal, this work has following limitations that
offer opportunities for future work.

\subsection{Explore Ways to Express or Assess Beliefs}
\system{} focuses on faithfully representing analysts’ beliefs through the dataset they construct via the interactive visualizations. 
However, to avoid missing values, only data points that have been fully connected across all variables (i.e., complete rows) are used when deriving statistical priors. 
\revision{
This requirement can make the process tedious, especially when  the number of variables increases and analysts must manually complete relationships for each variable. 
Future research should explore ways to reduce this burden. For example, analysts might ``sketch'' distributions or relationships. Alternatively, analysts could articulate their beliefs in natural language. An interesting technical challenge is in how to synthesize a coherent dataset from these multi-modal higher-level specifications, which could contradict each other.


Furthermore, \system{} does not assess the quality or validity of expressed beliefs. We focused on faithful representation because we suspected, and our user study confirmed, that analysts are often unsure whether their specifications in current prior elicitation tools reflect their intent. 
%
Future work should explore ways to assess the veracity of expressed beliefs and account for analysts’ meta-uncertainty~\cite{oakley2007uncertainty, hartmann2020prior}. For instance, systems could allow analysts to assign confidence weights \cite{ma2024you} to different parts of their specification and delegate the other unknown areas system to the system to fill in.
Alternatively, systems might leverage large language models’ reasoning capabilities and general knowledge to flag implausible assumptions or suggest revisions. 
Collectively, these directions point toward intelligent ``linters'' for Bayesian priors that support analysts in not only expressing their beliefs but also validating and refining them.
}



\subsection{Explore Alternative Methods for Deriving Statistical Priors}

Using the dataset that analysts construct through the interactive visualizations, \system{} repeatedly samples from the dataset and fits a previously specified statistical model. We developed this approach based on the definition of a prior as the set of ``reasonable'' possible values that a coefficient could take, given the analyst's beliefs.
Future work could explore alternative methods for deriving priors, such as simulations~\cite{bockting2024simulation} or Bayesian inference, that leverage the constructed dataset, histograms, or relationships..

\revision{
\system{} primarily encourages analysts to construct informative priors that reflect their expressed beliefs. Future research could investigate how to better support other types of priors, such as weakly informative priors or containment priors~\cite{Gelman2008weaklyprior, simpson2017penalising}.
%
One promising direction is to support prior specification in both the observable and parameter spaces~\cite{mikkola2024prior}, thereby balancing expressivity with fine-grained control. For example, after analysts articulate their domain knowledge in the observable space and obtain the derived priors, they could transition to a parameter-space view to further refine priors in detail. An interesting open question is how to design guardrails that ensure adjustments in either space preserve the previously expressed knowledge, or that surface divergences when more substantial revisions are needed. 
}

\subsection{Further Support Iterative Refinement of Priors and Models}
Iterative refinement of both priors and model structure is central to Bayesian analysis workflows~\cite{gelman2020workflow}, yet \system{} only supports iteration on priors via prior predictive checks and displays only the predictive distribution of the outcome variable.
Researchers should therefore explore mechanisms to allow analysts to refine both. 
\revision{
For example, future systems could involve exposing aspects of model specification to analysts, such as link functions, variable inclusion, or distribution families.
Future designs could also make the predictive-checking process more informative and transparent. For instance, by sampling jointly from the constructed dataset, overlaying predictive checks results on top of coordinated visualizations, or incorporating alternative visualizations such as hypothetical outcome plots (HOPs) to display the generated dataset~\cite{jessica2015hops}. Such extensions would help analysts more fully understand model behavior and uncertainty, and allow them to iterate on both model and priors as recommended in Bayesian workflow practice.
}

\subsection{Wider Range of Variables, Models and Users}
In this work, we scoped the statistical models supported in \system{} to
generalized linear models (without mixed effects) involving only continuous variables, and we only evaluated \system{} with Bayesian novices. Now that we have gathered evidence that viewing prior elicitation as an iterative dataset construction process is promising, future work should investigate how far this approach can generalize.

\revision{
One direction is extending \system{} to support categorical variables. This will require new forms of interactive visualizations, such as Sankey diagrams, to help analysts externalize their beliefs. 
Another direction is supporting more complex models, such as non-linear and mixed-effects models. This would enable the elicitation of a broader range of priors, including joint priors that encode dependencies or correlations among parameters.}

\revision{
Additionally, as the number of variables and model complexity increase, \system{}’s fixed layout and visualization choices (e.g., historgams, scatterplots) do not scale well and would quickly overwhelm analysts. Future versions could introduce flexible canvases, variable filtering, and on-demand visualization panels to better support high-dimensional data analyses. 
These extensions would also broaden the kinds of empirical studies \system{} can support, enabling researchers to examine prior elicitation practices with experienced Bayesian practitioners and in settings that more closely reflect real-world analysis workflows.
}

\section{Conclusion}
\system{} transforms prior elicitation into an iterative process of constructing a dataset that represents the knowledge of analysts. 
Through interactive visualizations, \system{} enables analysts to express their assumptions about observable variables, such as their distributions and relationships with other variables. These interactions construct an underlying dataset that \system{} uses to derive priors for a statistical model. 
In a controlled lab study with Bayesian novices, we found that \system{} lowered barriers to prior elicitation by helping participants externalize their knowledge and refine priors through actionable predictive feedback. Compared to trial-and-error approaches in the baseline, \system{} gave participants greater control, clarity, and confidence, leading to priors that better aligned with their expectations. 
Also, \system{} made Bayesian analysis feel approachable and increased participants’ willingness to use it in the future. These results suggest that interactive dataset construction is a promising step toward wider adoption of Bayesian analysis methods.


\section{Availability}

Source code and additional information are available at \url{https://github.com/ucla-cdl/prior-weaver}.

\begin{acks}
We thank the anonymous reviewers for their thoughtful feedback and our participants for their time and insights. We also thank members of the UCLA Computation \& Discovery Lab for feedback on the system and drafts of this paper, and Anna Riha for early discussions and feedback.
This work was supported by the ERC Advanced Grant (No. 101141916) and by the Research Council of Finland through the FCAI project (Grant Nos. 328400, 345604, 341763) and the Subjective Functions project (Grant No. 357578).
\end{acks}

\bibliographystyle{ACM-Reference-Format}
\bibliography{references}

\newpage
\onecolumn
\setcounter{page}{1}
\appendix
\section*{Appendix}
\section{Baseline Interface}
\begin{figure*}[h!]
    \centering
    \includegraphics[width=\linewidth]{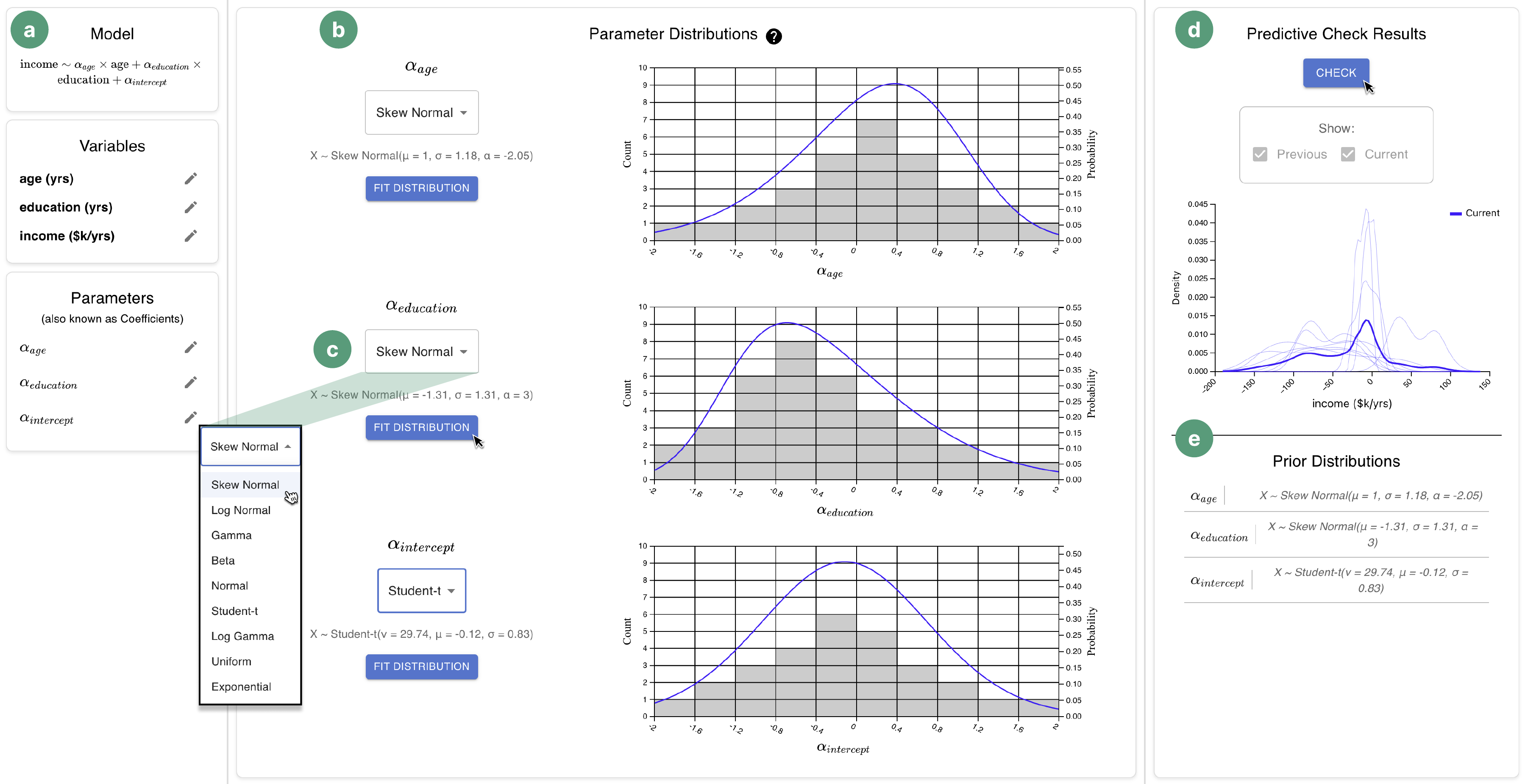}
    \caption{UI of the parameter space baseline: (a) an information panel displaying the model formula, variables, and parameters; (b) histograms where users can click and sketch parameter distributions; (c) after sketching, users can click \textsc{Fit Distribution} to obtain candidate continuous distributions matching their sketch and select the desired one; (d) users can click \textsc{Check} to receive feedback on expressed knowledge via prior predictive checks; and (e) the final prior distributions selected by users.}
    \label{fig:baseline}
    \Description[Screenshot of the parameter-space baseline interface.]{A five-part interface: (a) Information panel with model formula and variables; (b) Three histograms where users click to sketch parameter distributions; (c) Drop-down menus to select continuous distribution families (e.g., Skew Normal, Gamma) ; (d) A "CHECK" button and prior predictive check plot ; and (e) A text-based summary of selected prior distributions.}
\end{figure*}

\newpage

\section{Detailed Measurements}

Table \ref{tab:measurement} shows our detailed metrics and questions used in the user study.

\renewcommand{\arraystretch}{1.5}
\begin{table*}[h]
\centering  
\fontsize{8}{8}\selectfont  

\caption{Measurements used in our user study. For the survey items, a 5-point Likert scale was used, with 1 indicating ``Strongly disagree'' and 5 indicating ``Strongly agree''.}

\begin{tabular}{m{3cm}m{10cm}}
\hline
\textbf{Data Source} &
\textbf{Detailed Metric/Question} \\
\hline

\multirow{2}{*}{\makecell[l]{System log}} &
    Metrics of interaction patterns (e.g., number of link uses, generate uses, histogram edits) \\ \cline{2-2} 
    & 
    Number of iterations; predictive check results between iterations \\
\hline
\multirow{8}{*}{\makecell[l]{Survey}} &
    [\emph{Clear Expression}]: "I can clearly express my knowledge using this system." \\ \cline{2-2}
    &
    [\emph{Visualization Alignment w/ Knowledge}]: "I think the visualization results align with my knowledge." \\ \cline{2-2}
    &
    [\emph{Comfortable}]: "I feel comfortable expressing my knowledge using this system." \\ \cline{2-2}
    &
    [\emph{Ease of Expression}]: "This system makes it easy for me to express my knowledge." \\
    &
    [\emph{Translation Accuracy}"]: I believe my expressed knowledge is accurately translated into prior distributions." \\ \cline{2-2}
    &
    [\emph{Final Prior Alignment w/ Knowledge}]: "The final prior distributions I choose are aligned with my knowledge." \\ \cline{2-2}
    &
    [\emph{Understanding of Final Prior}"]: I understand the meaning of the finally elicited priors." \\ \cline{2-2}
    &
    [\emph{Becoming Appropriate}]: ``I think my expressed knowledge became more and more appropriate through the iterations.'' \\
\hline
\multirow{10}{*}{\makecell[l]{Survey \\ (System Usability Scale)}} &
    [\emph{Helpfulness}]: "Overall, I think this system is helpful in supporting prior elicitation." \\ \cline{2-2}
    &
    [\emph{Complexity}]: "I found the system unnecessarily complex." \\ \cline{2-2}
    &
    [\emph{Easy of Use}]: "I thought the system was easy to use." \\ \cline{2-2}
    &
    [\emph{Technical Support Needs}]: "I would need support of a technical person." \\ \cline{2-2}
    &
    [Function Integration]: "Functions are well integrated." \\ \cline{2-2}
    &
    [\emph{Quick to Learn}]: "I would imagine that most people would learn to use this system very quickly." \\ \cline{2-2}
    &
    [\emph{Confidence in Use}]: "I feel very confident using this system." \\ \cline{2-2}
    &
    [\emph{Cumbersome to Use}]: "I found the system cumbersome to use." \\ \cline{2-2}
    &
    [\emph{Learning Curve}]: "I needed to learn a lot before getting started." \\ \cline{2-2}
    &
    [\emph{Future Use}]: "I am inclined to use Bayesian analysis in my future experiments and analyses if I have access to this system." \\
\hline
\multirow{4}{*}{\makecell[l]{Interview}} &
  How do you express your knowledge using these visualizations? \\ \cline{2-2}
  &
  How do you decide that your expressed knowledge needs refinement, and what strategies do you use to do so? \\ \cline{2-2}
  &
  How would you compare your experiences with eliciting priors in the parameter space versus the observable space? \\  \cline{2-2}
  &
  Which of the two tools would make you more inclined to use Bayesian analysis in the future? Why? \\ 
\hline
  
\end{tabular}
\label{tab:measurement}
\end{table*}

\clearpage

\section{Detailed Evaluation Results}


\begin{figure*}[htbp]
    \centering
    \includegraphics[width=0.9\linewidth]{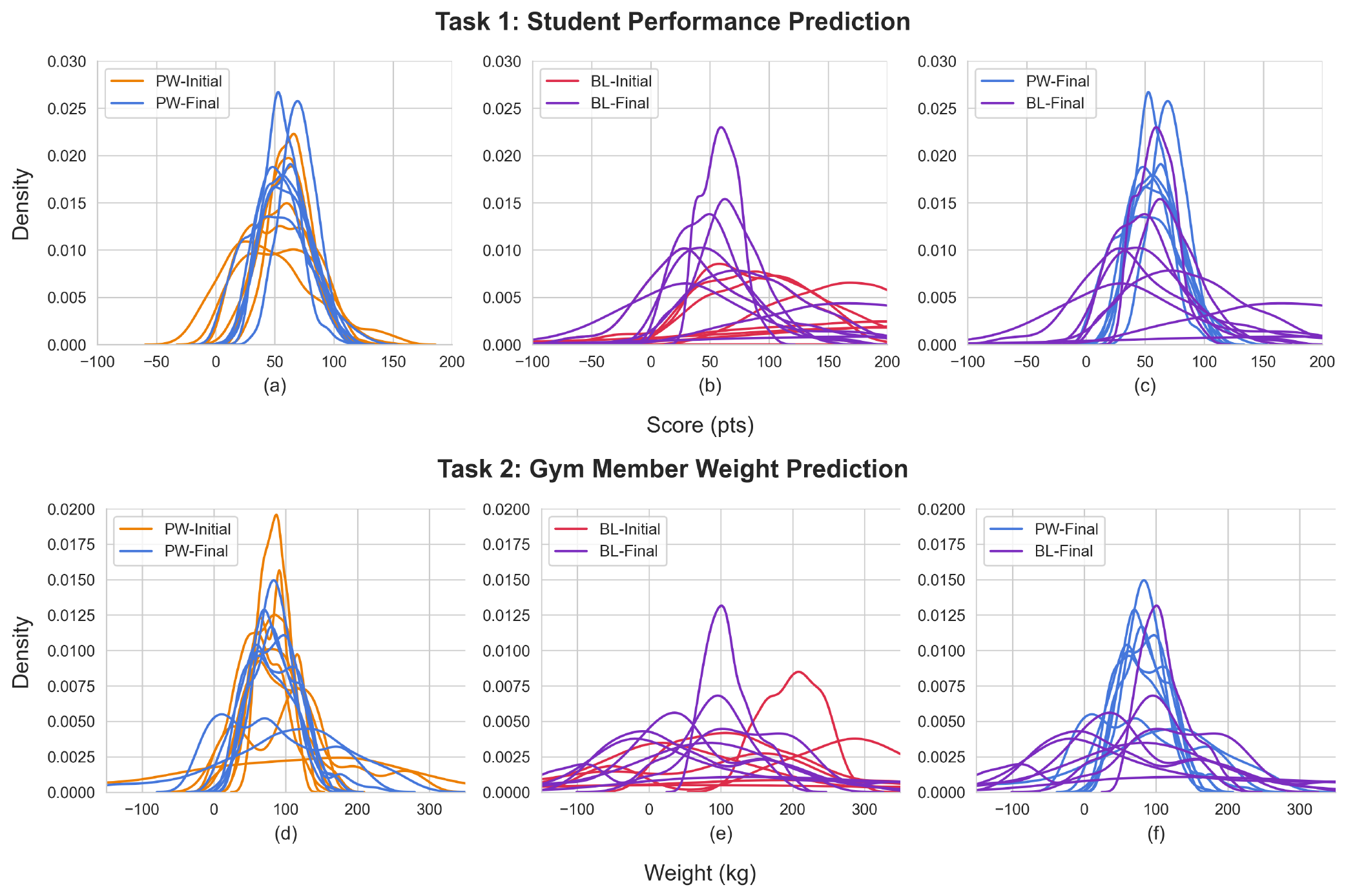}
    \caption{Predictive distributions of all participants across two tasks.
    \textit{PW} denotes \system{} and \textit{BL} denotes the baseline. 
    Figures (a) and (d) compare initial and final priors in \system{}.
    Figures (b) and (e) compare initial and final priors in the baseline. 
    Figures (c) and (f) compare the final priors between \system{} and the baseline.  
    }
    \Description[Density plots comparing initial and final predictive distributions for all participants.]{Six sub-plots (a–f) across two tasks. Charts (a) and (d) show \system{}'s initial and final distributions are stable and bell-shaped. Charts (b) and (e) show baseline distributions are initially very flat and undergo large, inconsistent shifts. Charts (c) and (f) highlight that \system{} results are more concentrated and plausible than the baseline's broad, less reasonable curves.}
    \label{fig:ppd-results-line}
\end{figure*}

\begin{figure}[htbp]
	\centering 
	\includegraphics[width=0.6\linewidth]{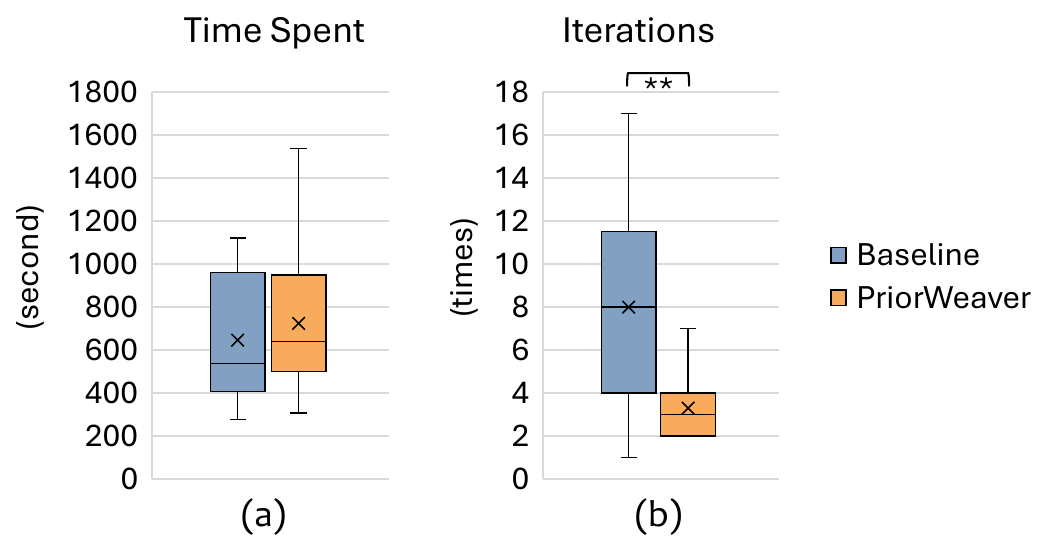}
	\caption{Participants’ engagement in the prior elicitation process with the two systems: (a) time spent across the process and (b) number of refinement iterations. These results suggest that although participants spent similar amounts of time in both conditions, the additional iterations in the baseline condition reflected trial-and-error adjustments rather than purposeful refinement of priors. (*: p < 0.05; **: p < 0.01; ***: p < 0.001).}
	\label{fig:findings:log-survey}
    \Description[Box plots showing time spent and number of iterations per condition.]{Two charts: (a) "Time Spent" box plot shows similar median times for both conditions, around 600–700 seconds. (b) "Iterations" box plot shows the baseline had a significantly higher number of refinement steps (median of 8) compared to \system{} (median of 3), with a statistical significance of $p < 0.01$.}
\end{figure}

\begin{figure*}[htbp]
	\centering 
	\includegraphics[width=\linewidth]{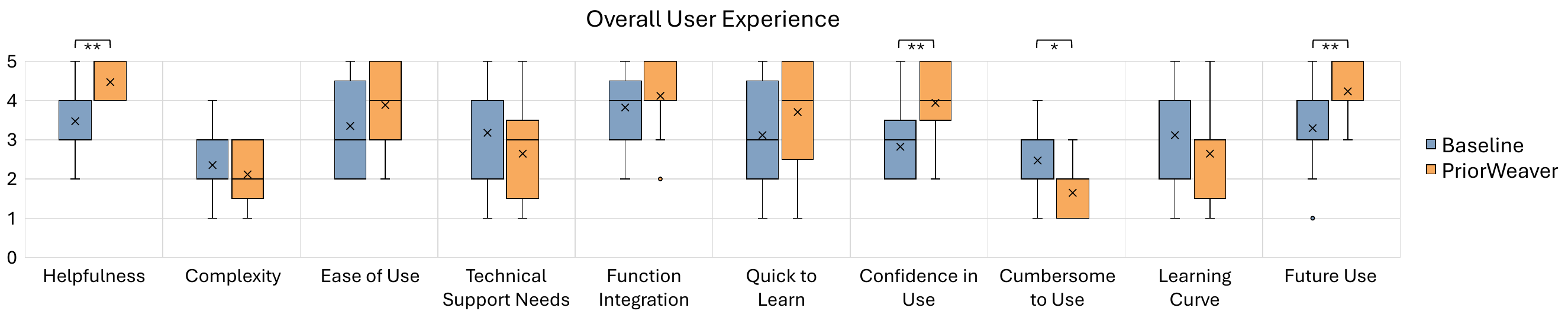}
	\caption{Survey results from the System Usability Scale (SUS). 
    Participants’ overall experience with the two systems indicated that they found \system{} more helpful, felt more confident using
	it, and were more inclined to use it for future Bayesian analysis. (*: p <
	0.05; **: p < 0.01; ***: p < 0.001).}
	\label{fig:findings:overallux-survey}
    \Description[Box plots of System Usability Scale (SUS) survey results.]{Displays 11 usability metrics on a 1-to-5 scale. \system{} scored significantly higher in Helpfulness, Confidence in Use, and Future Use ($p < 0.01$). \system{} also scored lower on "Cumbersome to Use" ($p < 0.05$), indicating a more positive user experience than the baseline.}
\end{figure*}

\end{document}